\pdfoutput=1
\documentclass[aps,prd,reprint,amsmath,amssymb,nofootinbib,longbibliography]{revtex4-2}

\usepackage[T1]{fontenc}
\usepackage{bm}
\usepackage{graphicx}
\usepackage{booktabs}
\usepackage{xcolor}
\usepackage{array}
\usepackage{tikz}
\usepackage{enumitem}
\usetikzlibrary{positioning,arrows.meta,fit,backgrounds,calc}
\usepackage{hyperref}
\usepackage{orcidlink}
\hypersetup{hidelinks}
\graphicspath{{./fig/}}

\definecolor{limc}{RGB}{193,66,66}
\definecolor{kunc}{RGB}{45,95,160}
\definecolor{lic}{RGB}{30,120,90}
\definecolor{lgray}{RGB}{240,240,242}

\newcommand{\LR}{\textcolor{limc}{\textbf{LR}}}
\newcommand{\KST}{\textcolor{kunc}{\textbf{KST}}}
\newcommand{\LI}{\textcolor{lic}{\textbf{LWYL}}}

\newcommand{\av}[1]{\left\langle #1\right\rangle}
\newcommand{\M}{\mathcal M}
\newcommand{\Hf}{\mathcal H}
\newcommand{\Kmq}{K_{\rm MQ}}
\newcommand{\nohyII}{\hyphenpenalty=10000\exhyphenpenalty=10000\relax}
\setlist{itemsep=1pt,parsep=0pt,topsep=2pt}
\newif\ifreprint\reprinttrue 
\ifreprint\newcommand{\admbreak}{}\else\newcommand{\admbreak}{\\&}\fi

\begin{document}

\title{Post-Newtonian secular dynamics of hierarchical triples.\ II.\\
Canonical derivation of the 1PN\,$\times$\,quadrupole cross terms\\
and of the quadrupole-squared term}

\author{Hideyoshi Arakida\,\orcidlink{0000-0002-0319-9305}}
\email{arakida.hideyoshi@nihon-u.ac.jp}
\affiliation{College of Engineering, Nihon University,
1 Nakagawara, Tokusada, Tamuramachi, Koriyama, Fukushima 963-8642, Japan}

\date{\today}

\begin{abstract}
We derive the secular Hamiltonian of a hierarchical triple from the 1PN ADM
three-body Hamiltonian by a two-stage Lie--Hori--Deprit transformation carried out
entirely in canonical variables, and use it to settle the relativistic cross terms
that were left open in Paper~I. All contributions to each order in the 1PN parameter
$\delta$ and the hierarchy parameter $\varepsilon$ are enumerated by a bookkeeping
rule for the two stages. The 1PN interaction of the monopole type reduces to a
renormalization of the mass of the inner binary by its binding energy, equal for the
inertial and the gravitational mass, and the order $\delta\varepsilon^{3/2}$, at which
Lim and Rodriguez (LR) found a libration term, is empty. The leading 1PN\,$\times$\,quadrupole
(electric) cross term and the magnetic-quadrupole cross term agree exactly with the
effective-field-theory result of Kuntz, Serra, and Trincherini; for the latter this
is the first independent derivation, and its mass factor splits as
$(4m+3m_3):m_3$ between the direct term and a second-stage cross term. We give the
next-to-leading order 1PN\,$\times$\,quadrupole term in closed form, verified by an
independent numerical construction; it depends on the outer argument of pericenter.
At quadrupole-squared order we obtain the complete secular Hamiltonian, including a
$2\omega_2$ harmonic; it is the Hamiltonian that generates the secular equations of
Will and of Conway and Will. The harmonic is a gauge term: the obstruction integral
that decides its removability vanishes even on the commensurability between the
von Zeipel--Lidov--Kozai (ZLK) period and the nodal precession, and the removing
generator is given in closed form as the shift of the first-order generator induced
by Tremaine's fictitious time. The EIH Lagrangian and the ADM Hamiltonian coincide at
1PN order; what differs between the formulations is the definition of the elements,
and we give the osculating--contact map including a term, induced by the tertiary,
of relative magnitude $3Gm_3/(c^2R)$. In secular integrations of the triples of LR, the relativistic
modulations and the eccentricity enhancement with the octupole are controlled by the
$\omega_2$-dependent part of the next-to-leading order term; with the canonical coefficient the
enhancement is two to three times that of Paper~I, and LR's accelerated mergers remain absent.
\end{abstract}

\maketitle

\section{Introduction}
\label{sec:intro}

The secular dynamics of hierarchical triples --- an inner binary of masses $m_1$
and $m_2$ perturbed by a distant tertiary of mass $m_3$ --- is governed at leading
order by the von Zeipel--Lidov--Kozai (ZLK) mechanism
\cite{vonZeipel1910,Lidov1962,Kozai1962,ItoOhtsuka2020,Naoz2016,Shevchenko2017}.
When the inner binary is compact, general relativity enters through the 1PN
precessions of the two orbits and through genuine three-body couplings, the
\emph{cross terms} of order $\delta\,\varepsilon^{m/2}$ in the 1PN parameter
$\delta=Gm/(c^2a_1)$ and the hierarchy parameter $\varepsilon=a_1/a_2$
\cite{NaozKocsisLoebYunes2013,Will2014,Will2014b,LimRodriguez2020,
KuntzSerraTrincherini2021,KuntzSerraTrincherini2023,LiWuYounsiLi2025}.
Relativistic effects in triples also matter observationally: the 1PN pericenter precession
shapes the response of an eccentric inner binary to a nearby tertiary, and hence the timing
of its gravitational-wave bursts, from which the mass and distance of the tertiary may be
inferred \cite{SainiEtAl2025}.

In the preceding paper \cite{PaperI} (Paper~I) we compared three formulations of
these cross terms: that of Lim and Rodriguez \cite{LimRodriguez2020} (\LR), based on
the Einstein--Infeld--Hoffmann (EIH) equations, osculating elements and a
two-parameter multiple-scale expansion; that of Kuntz, Serra, and Trincherini
\cite{KuntzSerraTrincherini2021,KuntzSerraTrincherini2023} (\KST), based on
nonrelativistic effective field theory and contact elements; and that of Li, Wu,
Younsi, and Li \cite{LiWuYounsiLi2025} (\LI), based on the ADM Hamiltonian and a
von Zeipel transformation. We traced LR's ``libration'' term at order
$\delta\varepsilon^{3/2}$ [their Eq.~(4.5)] to an inconsistent averaging measure and
showed that it vanishes with the time measure required of a canonical generating
function; we found that LR's term at order $\delta\varepsilon^{7/2}$ disagrees with
the magnetic-quadrupole term of KST in its mass dependence and angular structure and
fails a Hamiltonicity test. Paper~I left three questions open: a first-principles
canonical derivation of the $\delta\varepsilon^{7/2}$ term, which would provide an
independent check of KST; the remaining 1PN\,$\times$\,quadrupole cross terms at
orders $\delta\varepsilon^{3}$ and $\delta\varepsilon^{4}$; and the Newtonian
quadrupole-squared (Brown) term, at which the chain-rule defect of LR's
implementation identified in Paper~I becomes effective.

The Newtonian second-order problem has a history of its own. The secular Hamiltonian
at quadrupole-squared order, of relative magnitude
$\varepsilon_{21}\propto(n_2/n_1)(m_3/M)$ with respect to the ZLK Hamiltonian, goes
back to Brown \cite{Brown1936a,Brown1936b,Brown1936c} and has been rederived and
applied repeatedly
\cite{LuoKatzDong2016,LeiCirciOrtore2018,Tremaine2023,LeiGrishin2025a,LeiGrishin2025b}.
Will \cite{Will2021} and Conway and Will \cite{ConwayWill2024} obtained the
corresponding secular equations by a multiple-scale method and found terms that
depend on the outer argument of pericenter through $\cos(2\omega_1\pm2\omega_2)$,
which are absent from the Brown Hamiltonian in the form of Refs.~\cite{Tremaine2023,
LeiGrishin2025a}. Tremaine \cite{Tremaine2023} showed that the form of the Brown
Hamiltonian depends on the choice of a fictitious time used in the average over the
outer orbit, and that the $\omega_2$-dependent part can be made to vanish. Whether
this freedom holds everywhere, in particular where the ZLK period is commensurate
with the nodal precession, has not been examined.

In this paper we derive all these terms within a single canonical framework. We
start from the 1PN ADM Hamiltonian of three bodies in Jacobi coordinates
\cite{OhtaEtAl1974,Schafer1987} and eliminate the two mean anomalies by a two-stage
Lie--Hori--Deprit transformation \cite{Hori1966,Deprit1969}, with the time measure
throughout. Canonical variables are kept from the beginning to the end, so that the
contributions that were distributed over several ``channels'' in the multiple-scale
calculation (Paper~I) appear as the expansion terms of single Poisson brackets with the
generator of the transformation (Sec.~\ref{sec:lie}), and the
result is a secular Hamiltonian by construction. A bookkeeping rule for the orders
generated at each stage allows us to enumerate every contribution to a given order,
direct and indirect, and to prove that the others vanish. Our main results are:
\begin{enumerate}[label=(\roman*),leftmargin=16pt]
\item The 1PN monopole interaction is a renormalization of the mass of the inner
      binary by its binding energy, equal for the inertial and gravitational mass
      (Sec.~\ref{sec:mono}); the order $\delta\varepsilon^{3/2}$ is empty, in the
      Hamiltonian itself and in the cross terms.
\item The leading (electric) 1PN\,$\times$\,quadrupole cross term agrees exactly with
      KST, for all eccentricities and masses (Sec.~\ref{sec:pnq-e}).
\item The magnetic-quadrupole cross term agrees exactly with KST; this is its first
      independent derivation. Its mass factor $m_3(4m+3m_3)$ is shared between the
      direct term and a second-stage cross term in the ratio $(4m+3m_3):m_3$
      (Sec.~\ref{sec:pnq-m}).
\item The next-to-leading order 1PN\,$\times$\,quadrupole cross term is obtained in closed
      form and verified by an independent numerical construction
      (Sec.~\ref{sec:pnq-n}). It depends on $\omega_2$ through the outer
      eccentricity, so that the component of the inner angular momentum normal to
      the outer orbit is not conserved at this order.
\item At quadrupole-squared order we obtain the complete secular Hamiltonian in
      closed form, including a $2\omega_2$ harmonic. It is the Hamiltonian that
      generates the secular equations of Will \cite{Will2021} and of Conway and Will
      \cite{ConwayWill2024} (Sec.~\ref{sec:q2}). The harmonic is a gauge term: the
      question reduces to two first-order partial differential equations whose
      characteristics are the ZLK orbits, and the obstruction integral vanishes even on
      the commensurability $\Delta\Omega_1=k\pi$; the removing generator is the shift
      of the first-order generator induced by Tremaine's fictitious time, which we give in
      closed form.
\item The EIH Lagrangian and the ADM Hamiltonian give the same 1PN Hamiltonian; what
      distinguishes the formulations is the definition of the orbital elements
      and of the mean elements (Sec.~\ref{sec:ham-eih}). The map between osculating
      and contact elements contains, besides the two-body part, a term of relative magnitude
      $3Gm_3/(c^2R)$ induced by the tertiary.
\end{enumerate}
We then compare the orbital evolution given by LR's secular equations, by those of
Paper~I, and by the canonical secular Hamiltonian of this paper for the triples used
by LR (Sec.~\ref{sec:orbit}). The recomputation of LR's population of triples is
the subject of Paper~III.

The paper is organized as follows. Section~\ref{sec:ham} introduces the Hamiltonian,
the relation between the EIH and ADM formulations, and the ordering.
Section~\ref{sec:lie} sets up the canonical perturbation theory, the gauge theorem and
the enumeration of the contributions. Sections~\ref{sec:mono}--\ref{sec:q2} derive the
monopole and dipole terms, the 1PN\,$\times$\,quadrupole cross terms and the
quadrupole-squared term. Section~\ref{sec:orbit} compares the orbital evolution,
Sec.~\ref{sec:disc} discusses the results in relation to previous work, and
Sec.~\ref{sec:sum} summarizes. Appendix~\ref{app:vz} compares the von Zeipel and
Lie--Hori--Deprit methods, Appendix~\ref{app:kepler} collects Kepler averages,
Appendix~\ref{app:pipe} describes the symbolic and numerical pipelines, and
Appendix~\ref{app:expr} lists intermediate expressions. We use the notation of
Paper~I; the names of the secular terms and their orders are collected in
Table~\ref{tab:names}.

\section{Hamiltonian, elements and orders}
\label{sec:ham}

\subsection{The 1PN three-body Hamiltonian}

\begin{figure*}[tb]
\centering
\resizebox{\linewidth}{!}{\input{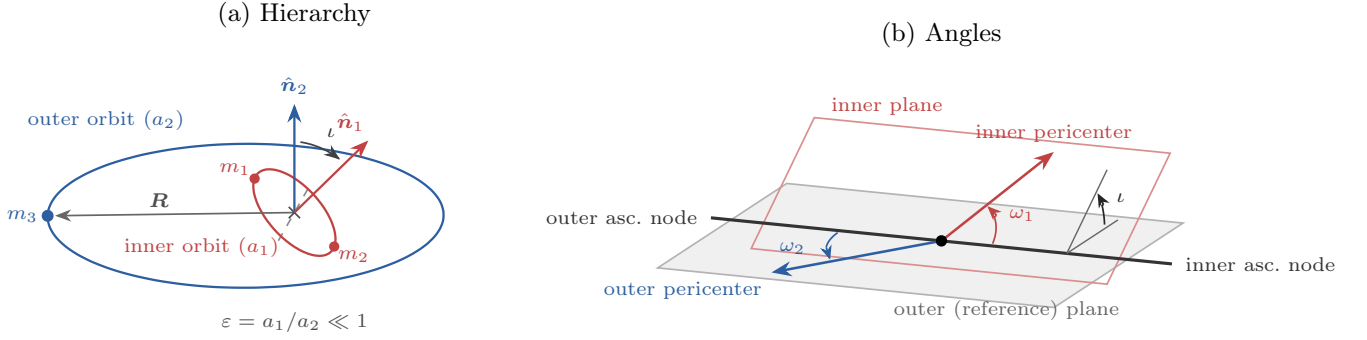}}
\caption{Geometry of the hierarchical triple (as in Fig.~1 of Paper~I).
(a) The inner binary $(m_1,m_2)$ with semimajor axis $a_1$ and the tertiary $m_3$
with semimajor axis $a_2$. The outer Jacobi vector $\bm R$ points from the center of
mass of the inner binary (cross) to $m_3$; the inner Jacobi vector is
$\bm r=\bm x_1-\bm x_2$, and the hierarchy parameter is $\varepsilon=a_1/a_2$. The
unit normals $\hat{\bm n}_1$ and $\hat{\bm n}_2$ (directions of the orbital angular
momenta $\bm G_1$ and $\bm G_2$) make the mutual inclination $\iota$,
Eq.~\eqref{eq:cosi}; the dashed line is the line of nodes.
(b) Oblique view of the angles. The reference plane is the invariable plane (normal
to the total angular momentum $\bm J$), which nearly coincides with the outer orbital
plane (gray). The ascending nodes of the two orbits lie at opposite ends of the node
line, $\Omega_2=\Omega_1+\pi$ (node elimination, Sec.~\ref{sec:ham-del}). The
arguments of pericenter $\omega_1$ and $\omega_2$ are measured in the inner and outer
planes from the respective ascending nodes, in the direction of motion; together with
the mean anomalies $\M_1,\M_2$ they are the angles of the Delaunay variables.}
\label{fig:geom}
\end{figure*}

We use the notation of Paper~I: $m=m_1+m_2$, $M=m+m_3$, $\mu_1=m_1m_2/m$,
$\mu_2=mm_3/M$, $\nu=m_1m_2/m^2$, the inner and outer Jacobi vectors
$\bm r=\bm x_1-\bm x_2$ and $\bm R=\bm x_3-(m_1\bm x_1+m_2\bm x_2)/m$, and
$\ell_i=\sqrt{1-e_i^2}$; the geometry is shown in Fig.~\ref{fig:geom}. In the barycentric frame the canonical momenta conjugate to
$\bm r$ and $\bm R$ are $\bm p$ and $\bm P$, with
\begin{equation}
\bm p_1=\bm p-\frac{m_1}{m}\bm P,\qquad
\bm p_2=-\bm p-\frac{m_2}{m}\bm P,\qquad
\bm p_3=\bm P .
\label{eq:jacp}
\end{equation}

The Hamiltonian is $H=H_{\rm N}+c^{-2}H_{\rm 1PN}$, with the ADM 1PN part
\cite{OhtaEtAl1974,Schafer1987}
\begin{widetext}
\begin{equation}
\begin{split}
H_{\rm 1PN}={}&-\frac18\sum_a\frac{(p_a^2)^2}{m_a^3}
-\frac{G}{4}\sum_{a\ne b}\frac{m_am_b}{r_{ab}}
\left[6\frac{p_a^2}{m_a^2}-7\frac{\bm p_a\!\cdot\!\bm p_b}{m_am_b}
-\frac{(\bm n_{ab}\!\cdot\!\bm p_a)(\bm n_{ab}\!\cdot\!\bm p_b)}{m_am_b}\right]
\admbreak
+\frac{G^2}{2}\sum_a\sum_{b\ne a}\sum_{c\ne a}\frac{m_am_bm_c}{r_{ab}r_{ac}},
\end{split}
\label{eq:adm}
\end{equation}
\end{widetext}
where the last sum includes $b=c$. The Newtonian part is
\begin{equation}
\begin{split}
H_{\rm N}={}&\frac{p^2}{2\mu_1}-\frac{Gm\mu_1}{r}+\frac{P^2}{2\mu_2}-\frac{GM\mu_2}{R}\\
&-\frac{Gm_3}{R}\sum_{k\ge2}\Big(\frac{r}{R}\Big)^k\mathcal M_k\,
P_k(\hat{\bm n}\!\cdot\!\hat{\bm N}),
\end{split}
\label{eq:HN}
\end{equation}
with
\begin{equation}
\begin{split}
\mathcal M_k&=m_1\Big(\frac{m_2}{m}\Big)^k+m_2\Big(-\frac{m_1}{m}\Big)^k\\
&=\frac{m_1m_2\big[m_2^{k-1}-(-m_1)^{k-1}\big]}{m^k};
\end{split}
\label{eq:Mk}
\end{equation}
the Newtonian dipole ($k=1$)
vanishes identically in Jacobi coordinates, the quadrupole ($k=2$) generates the ZLK
dynamics and the octupole ($k=3$) is proportional to $m_1m_2(m_1-m_2)$.
Eq.~\eqref{eq:HN} is the Jacobi-coordinate Legendre expansion of the stellar
three-body problem given by Harrington \cite{Harrington1969}: with his labels
$(m_0,m_1,m_2)$ for the inner pair and the tertiary, i.e.\ $(m_0,m_1,m_2)\to(m_2,m_1,m_3)$ here and his inner vector
pointing from $m_0$ to $m_1$, his mass factor
$M_j=m_0m_1m_2\,[m_0^{j-1}-(-m_1)^{j-1}]/(m_0+m_1)^j$ equals $m_3\mathcal M_k$ of
Eq.~\eqref{eq:Mk} at $j=k$, so that the interaction terms of the two expansions coincide
term by term. The same expansion underlies the Newtonian part of Paper~I and of
Refs.~\cite{NaozEtAl2013,Naoz2016}.

\subsection{EIH equations and ADM Hamiltonian}
\label{sec:ham-eih}

LR (and Paper~I) start from the EIH equations of motion in harmonic coordinates, the
present paper from the ADM Hamiltonian. At 1PN order the two are the same dynamical
system. The EIH Lagrangian of $N$ bodies \cite{EinsteinInfeldHoffmann1938,
LandauLifshitz1975} is
\begin{equation}
L=\sum_a\tfrac12m_av_a^2+\sum_{a<b}\frac{Gm_am_b}{r_{ab}}+c^{-2}L_{\rm 1PN}
\label{eq:LEIH0}
\end{equation}
with
\begin{equation}
\begin{split}
L_{\rm 1PN}={}&\sum_a\frac{m_av_a^4}{8}
+\frac12\sum_{a\ne b}\frac{Gm_am_b}{r_{ab}}\Big[3v_a^2-\frac72\bm v_a\!\cdot\!\bm v_b\\
&-\frac12(\bm n_{ab}\!\cdot\!\bm v_a)(\bm n_{ab}\!\cdot\!\bm v_b)\Big]\\
&-\frac12\sum_a\sum_{b\ne a}\sum_{c\ne a}\frac{G^2m_am_bm_c}{r_{ab}r_{ac}} .
\end{split}
\label{eq:LEIH}
\end{equation}
Since the 1PN term is a perturbation, its Legendre transform is
$H_{\rm 1PN}(\bm x,\bm p)=-L_{\rm 1PN}(\bm x,\bm v=\bm p/m)$ to this order, and
substituting $\bm v_a=\bm p_a/m_a$ in Eq.~\eqref{eq:LEIH} reproduces
Eq.~\eqref{eq:adm} term by term; we have also verified the identity symbolically for
three bodies. The harmonic and ADM coordinates therefore coincide at 1PN order; they
first differ at 2PN \cite{DamourSchafer1985}. The ``gauge'' differences between the
formulations are not differences of coordinates but of the variables in which the
secular dynamics is expressed. There are three of them.

\paragraph*{(i) Osculating versus contact elements.}
LR's elements are the Kepler elements of $(\bm r,\bm v)$ with $\bm v=\dot{\bm r}$
(osculating elements, labeled by the superscript ``osc''), whereas the canonical Delaunay
variables are the Kepler elements of $(\bm r,\bm p/\mu_1)$ (contact elements, superscript
``con'') \cite{EfroimskyGoldreich2003,Efroimsky2005}. Both sets are computed from the same
phase-space point with the same Kepler formulas; they differ only in whether the velocity or
the momentum divided by the reduced mass is inserted.
Their difference is fixed by
$\bm v-\bm p/\mu_1=\partial H_{\rm 1PN}/\partial\bm p/c^2$ (Paper~I, Appendix~B). The two-body part gives, averaged over the inner orbit,
\begin{align}
\av{\frac1{a_1^{\rm osc}}}-\frac1{a_1^{\rm con}}
&=\frac{Gm}{c^2a_1^2}\Big[\frac{16-6\nu}{\ell_1}-9+5\nu\Big],
\label{eq:oscon-a}\\
\av{e_1^{\rm osc}}-e_1^{\rm con}
&=-\frac{Gm}{c^2a_1}\,(8-3\nu)\,\frac{\ell_1(1-\ell_1)}{e_1},
\label{eq:oscon-e}
\end{align}
where $a_1^{\rm osc}$, $e_1^{\rm osc}$ are the osculating and $a_1^{\rm con}$, $e_1^{\rm con}$
the contact semimajor axis and eccentricity, the average is over the inner orbit, and the
second line is the component of the eccentricity vector along the pericenter; the
transverse component vanishes. Paper~I stated that this map depends on the inner orbit only. That
is true of the two-body part, but the tertiary also contributes. The terms of
Eq.~\eqref{eq:adm} that couple $p^2$ to $1/r_{13}$ and $1/r_{23}$ give, at monopole
order, $\bm v-\bm p/\mu_1\supset-(3Gm_3/c^2R)\,\bm p/\mu_1$: the velocity is
rescaled by the potential of the tertiary. Hence
\begin{gather}
\frac{a_1^{\rm osc}-a_1^{\rm con}}{a_1}\bigg|_{m_3}=-\frac{6Gm_3}{c^2R},\nonumber\\
\frac{G_1^{\rm osc}-G_1^{\rm con}}{G_1}\bigg|_{m_3}=-\frac{3Gm_3}{c^2R},
\label{eq:oscon-m3}
\end{gather}
with no change of $e_1$ (all symbolic, Appendix~\ref{app:pipe}). This term is of
order $\delta\varepsilon\,(m_3/m)$ and follows the outer orbit. For the triple of LR's Fig.~8, $m_3/m=4\times10^5$ and $Gm_3/(c^2a_2)\simeq1.4\times10^{-3}$, so that the osculating
semimajor axis of the inner orbit oscillates by about $4\%$ over an outer orbit with
$e_2=0.8$, while the contact $a_1$ does not. Through Eq.~(4) of Paper~I, a
near-identity map of relative order $\delta\varepsilon$ changes the secular rates at
order $\delta\varepsilon\times\varepsilon^3=\delta\varepsilon^4$, i.e., at the order of
the next-to-leading order 1PN\,$\times$\,quadrupole term. A term-by-term comparison of the
secular rates at that order is therefore meaningful only after the elements have been
converted (Sec.~\ref{sec:pnq-n}).

\paragraph*{(ii) The averaging measure.}
The mean elements depend on the normalization of the periodic part of the generator.
The canonical choice is time-mean-free; LR's periodic solutions are
mean-free with respect to the outer true anomaly. Paper~I showed that this
difference generates LR's Eq.~(4.5) at order $\delta\varepsilon^{3/2}$; the same
freedom reappears in Sec.~\ref{sec:q2} as the fictitious time of Tremaine.

\paragraph*{(iii) The center of mass.}
The Jacobi vector $\bm R$ is referred to the Newtonian center of mass of the inner
binary. Its 1PN correction is not a separate choice here: it is contained in the
odd-in-$\bm r$ part of the Hamiltonian at order $\delta\varepsilon^2$, and its effect on
the secular dynamics appears as the second-stage cross term of the magnetic-quadrupole
term (Sec.~\ref{sec:pnq-m}), the canonical counterpart of the center-of-mass issue
discussed by KST.

\subsection{Orders and names of the secular terms}
\label{sec:ham-orders}

With the inner binding energy as the unit, the Hamiltonian is ordered by
\begin{equation}
\delta=\frac{Gm}{c^2a_1},\quad \varepsilon=\frac{a_1}{a_2},\quad
\frac{V^2}{v^2}\sim\frac{M}{m}\varepsilon,\quad
\frac{Gm_3/R}{Gm/r}\sim\frac{m_3}{m}\varepsilon,
\label{eq:scal}
\end{equation}
where $v$ and $V$ are the inner and outer orbital velocities, and mass ratios are not
counted. A term scaling as $\delta^n\varepsilon^{m/2}$ is said to be of order
$(n,m/2)$ in the notation of Paper~I. In practice we set $\bm P\to\tau\bm P$ and
$\bm R\to\bm R/\tau^2$, so that the power $\tau^m$ labels $\varepsilon^{m/2}$.
Table~\ref{tab:names} lists the terms that appear below with the names we use for them.

\begin{table}[tb]
\caption{The secular terms treated in this paper, their scaling (relative to the
inner Kepler energy), the order label of Paper~I, and the section in which they are
derived.}
\label{tab:names}
\centering\small
\begin{tabular}{@{}p{0.47\columnwidth}ccc@{}}
\toprule
Term & Scaling & Order & Sec.\\
\midrule
ZLK (quadrupole) & $\varepsilon^3$ & $(0,3)$ & --\\
Octupole & $\varepsilon^4$ & $(0,4)$ & --\\
Quadrupole-squared (Brown) & $\varepsilon^{9/2}$ & $(0,9/2)$ & \ref{sec:q2}\\
\midrule
Inner 1PN precession & $\delta$ & $(1,0)$ & \ref{sec:mono-10}\\
Monopole, odd in $\bm P$ & $\delta\varepsilon^{1/2}$ & $(1,\tfrac12)$ & \ref{sec:mono-half}\\
Binding-energy (mass) renormalization & $\delta\varepsilon$ & $(1,1)$ & \ref{sec:mono-11}\\
LR libration term (absent) & $\delta\varepsilon^{3/2}$ & $(1,\tfrac32)$ & \ref{sec:mono-32}\\
Outer 1PN, 1PN dipole & $\delta\varepsilon^{2}$ & $(1,2)$ & \ref{sec:mono-2}\\
de Sitter (dipole\,$\times$\,velocity) & $\delta\varepsilon^{5/2}$ & $(1,\tfrac52)$ & \ref{sec:mono-dS}\\
\midrule
1PN\,$\times$\,quadrupole, electric & $\delta\varepsilon^{3}$ & $(1,3)$ & \ref{sec:pnq-e}\\
1PN\,$\times$\,quadrupole, magnetic & $\delta\varepsilon^{7/2}$ & $(1,\tfrac72)$ & \ref{sec:pnq-m}\\
1PN\,$\times$\,quadrupole, next-to-leading order & $\delta\varepsilon^{4}$ & $(1,4)$ & \ref{sec:pnq-n}\\
\bottomrule
\end{tabular}
\end{table}

\subsection{Delaunay variables and node elimination}
\label{sec:ham-del}

The Delaunay variables of the inner and outer orbits are $(\M_i,\omega_i,\Omega_i;
L_i,G_i,H_i)$, with $L_1=\mu_1\sqrt{Gma_1}$, $L_2=\mu_2\sqrt{GMa_2}$, $G_i=L_i\ell_i$ and
$H_i=G_i\cos\iota_i$ (Paper~I). With the reference plane normal to the total angular
momentum the ascending nodes are opposite, $\Omega_2=\Omega_1+\pi$, and the
Hamiltonian depends on the nodes only through the mutual inclination $\iota$,
\begin{equation}
\cos\iota=\frac{J^2-G_1^2-G_2^2}{2G_1G_2},
\label{eq:cosi}
\end{equation}
where $J$ is the total angular momentum. After node elimination the canonical pairs
are $(\M_1,L_1)$, $(\omega_1,G_1)$, $(\M_2,L_2)$ and $(\omega_2,G_2)$, and
\begin{equation}
\frac{\partial\cos\iota}{\partial G_1}=-\frac{1}{G_2}-\frac{\cos\iota}{G_1},\qquad
\frac{\partial\cos\iota}{\partial G_2}=-\frac{1}{G_1}-\frac{\cos\iota}{G_2}.
\label{eq:dcosi}
\end{equation}
The terms $1/G_2$ are of relative order $G_1/G_2\sim\varepsilon^{1/2}(m/m_3)^{1/2}$; they are
neglected in the test-particle frame, in which the outer orbit is fixed, $H_1=G_1\cos\iota$
is the momentum conjugate to $\Omega_1$ and $\partial/\partial\Omega_1=-\partial/\partial\omega_2$
on functions of $\omega_2=\varpi_2-\Omega_1$ (up to a constant). We use the exact pairs
where the $1/G_2$ terms contribute (Sec.~\ref{sec:pnq-n}) and the test-particle frame
otherwise.

The node elimination is carried out on the Hamiltonian, not on the equations of motion. The
Hamiltonian is invariant under rotations, so that, with the total angular momentum along the
reference axis, $\Omega_1$ and $\Omega_2$ enter only through $\Omega_2-\Omega_1=\pi$ and
$\cos\iota$ is the function of $(G_1,G_2,J)$ given by Eq.~\eqref{eq:cosi}. This is Jacobi's reduction of
the rotational symmetry: the reduced Hamiltonian $K(\omega_1,G_1,\omega_2,G_2;J)$ generates
the correct equations for the pairs $(\omega_i,G_i)$, provided $\cos\iota$ is differentiated
at fixed $J$, which is what Eq.~\eqref{eq:dcosi} does. The remaining variables are not
obtained from the reduced Hamiltonian: $H_1$ and $H_2$ follow from $\bm G_1+\bm G_2=\bm J$,
and the nodes from the precession of the line of nodes about $\bm J$. Deriving $\dot H_i$ from the
reduced Hamiltonian instead leads to the error at octupole order identified by Naoz et
al.~\cite{NaozEtAl2013} in the equations of Ref.~\cite{FordKozinskyRasio2000}. In the orbit integrations of
Sec.~\ref{sec:orbit} no nodes are eliminated: the code evolves the angular-momentum and
eccentricity vectors of both orbits, and the reduced Hamiltonian is only used to derive the
secular terms.

\subsection{Averaging measure}

Secular quantities are time averages over the mean anomalies,
$\av{\cdot}_{\M_i}=(2\pi)^{-1}\oint(\cdot)\,d\M_i$. In practice we write the inner orbit
in the eccentric anomaly $E$ and the outer orbit in the true anomaly $F$,
\begin{gather}
d\M_1=(1-e_1\cos E)\,dE,\nonumber\\
d\M_2=\frac{\ell_2^3}{(1+e_2\cos F)^2}\,dF\equiv w_2\,dF,
\label{eq:measure}
\end{gather}
in which the integrands become trigonometric polynomials
(Appendix~\ref{app:pipe}). The lesson of Paper~I is that the weight $w_2$ may not be
replaced by unity; in the canonical theory this is automatic, because the homological
equation, Eq.~\eqref{eq:homo} below, fixes the measure.

\section{Canonical perturbation theory}
\label{sec:lie}

\begin{figure*}[tb]
\centering
\resizebox{\linewidth}{!}{\begin{tikzpicture}[>=Stealth,line width=0.6pt,font=\small,x=1cm,y=1cm,
 bx/.style={draw=black!55,rounded corners=2pt,align=center,execute at begin node=\nohyII,
       inner sep=4pt,font=\small,minimum height=11mm,text width=41mm,fill=white},
 st/.style={bx,draw=kunc,line width=1.0pt,fill=kunc!6},
 res/.style={bx,draw=lic,line width=1.0pt,fill=lic!6},
 term/.style={bx,text width=64mm,minimum height=9mm,font=\scriptsize},
 aux/.style={draw=black!40,dashed,rounded corners=2pt,align=center,execute at begin node=\nohyII,
       inner sep=3pt,font=\scriptsize,minimum height=9mm,text width=41mm,fill=black!2},
 ar/.style={->,black!65},
 au/.style={->,black!40,dashed}]

\node[bx] (ham) at (0,0) {1PN ADM Hamiltonian\\[-1pt]\scriptsize Jacobi coordinates,
  Eqs.~\eqref{eq:adm}--\eqref{eq:HN}};
\node[bx] (ord) at (5.3,0) {order bookkeeping\\[-1pt]\scriptsize grades $\delta^n\varepsilon^{m/2}$,
  Table~\ref{tab:names}};
\node[st] (s1) at (10.6,0) {1st stage: eliminate $\M_1$\\[-1pt]\scriptsize $W\propto1/n_1$; inner Delaunay
  and outer $(\bm R,\bm P)$ brackets};
\node[st] (s2) at (15.9,0) {2nd stage: eliminate $\M_2$\\[-1pt]\scriptsize $W\propto1/n_2$;
  renormalized outer Kepler};

\node[aux] (eih) at (0,-2.0) {EIH (harmonic) $\equiv$ ADM at 1PN;\\
  osculating $\leftrightarrow$ contact elements (Sec.~\ref{sec:ham-eih})};
\node[aux] (tr) at (5.3,-2.0) {transfer of orders (Table~\ref{tab:transfer});\\
  complete enumeration (Table~\ref{tab:enum})};
\node[aux] (lhd) at (10.6,-2.0) {Lie--Hori--Deprit, Eqs.~\eqref{eq:K1}--\eqref{eq:K2avg};\\
  cross terms $a\otimes b$, Eq.~\eqref{eq:cross}};
\node[aux] (gau) at (15.9,-2.0) {time measure, gauge theorem, Eq.~\eqref{eq:gaugethm};\\
  $F$ measure $\Rightarrow$ LR Eq.~(4.5)};

\node[res] (K) at (15.9,-4.6) {double-averaged secular\\[-1pt] Hamiltonian $K$\\[-1pt]
  \scriptsize in mean Delaunay elements};
\node[term] (mono) at (9.0,-3.6) {inner 1PN, monopole, binding-energy renormalization,\\
  outer 1PN $+$ 1PN dipole, de Sitter (Sec.~\ref{sec:mono})};
\node[term] (pnq) at (9.0,-4.75) {1PN$\times$quadrupole: electric, magnetic-quadrupole,\\
  next-to-leading order (Sec.~\ref{sec:pnq})};
\node[term] (q2) at (9.0,-5.9) {quadrupole-squared: closed form; the $2\omega_2$ harmonic\\
  is a gauge term, removed by $\gamma_\tau$ (Sec.~\ref{sec:q2})};
\node[res] (orb) at (1.2,-4.75) {secular equations and\\[-1pt] orbit evolution\\[-1pt]\scriptsize (Sec.~\ref{sec:orbit})};

\node[aux,text width=120mm] (chk) at (9.0,-7.55) {checks: every term constructed symbolically (Mathematica) and by an
  independent numerical (FFT) construction (App.~\ref{app:pipe});
  comparison with KST, LWYL, LR, Will, and Conway--Will (Secs.~\ref{sec:pnq}, \ref{sec:q2}, \ref{sec:disc})};

\draw[ar] (ham) -- (ord);
\draw[ar] (ord) -- (s1);
\draw[ar] (s1) -- (s2);
\draw[ar] (s2.east) -- ++(0.45,0) |- (K.east);
\foreach \b in {mono,pnq,q2}{
  \draw[ar] (K.west) -- (13.15,-4.6) |- (\b.east);
  \draw[ar] (\b.west) -- (4.85,0|-\b.west) |- (orb.east);}
\draw[au] (eih) -- (ham);
\draw[au] (tr) -- (ord);
\draw[au] (lhd) -- (s1);
\draw[au] (gau) -- (s2);
\draw[au] (chk.north) -- (q2.south);
\end{tikzpicture}}
\caption{Calculation pipeline of this paper (compare Fig.~3 of Paper~I for the
multiple-scale pipeline of LR). The 1PN ADM Hamiltonian in Jacobi coordinates is
ordered by the grades $\delta^n\varepsilon^{m/2}$ and transformed in two
Lie--Hori--Deprit stages (blue), which eliminate the inner and the outer mean anomaly
with the time measure. The dashed boxes give the tools used at each step: the
equivalence of the EIH and ADM formulations and the map between osculating and
contact elements (Sec.~\ref{sec:ham-eih}), the transfer rules and the complete
enumeration of contributions (Tables~\ref{tab:transfer} and \ref{tab:enum}), the
second-order Lie--Hori--Deprit formulas with the cross terms $a\otimes b$, and the
gauge theorem that fixes the averaging measure. The result (green) is a secular
Hamiltonian in mean Delaunay elements whose named terms are derived in
Secs.~\ref{sec:mono}--\ref{sec:q2} and used for the orbit integrations of
Sec.~\ref{sec:orbit}. Every term is constructed twice, symbolically and by an
independent numerical (FFT) construction (Appendix~\ref{app:pipe}), and compared with
previous work.}
\label{fig:flow}
\end{figure*}
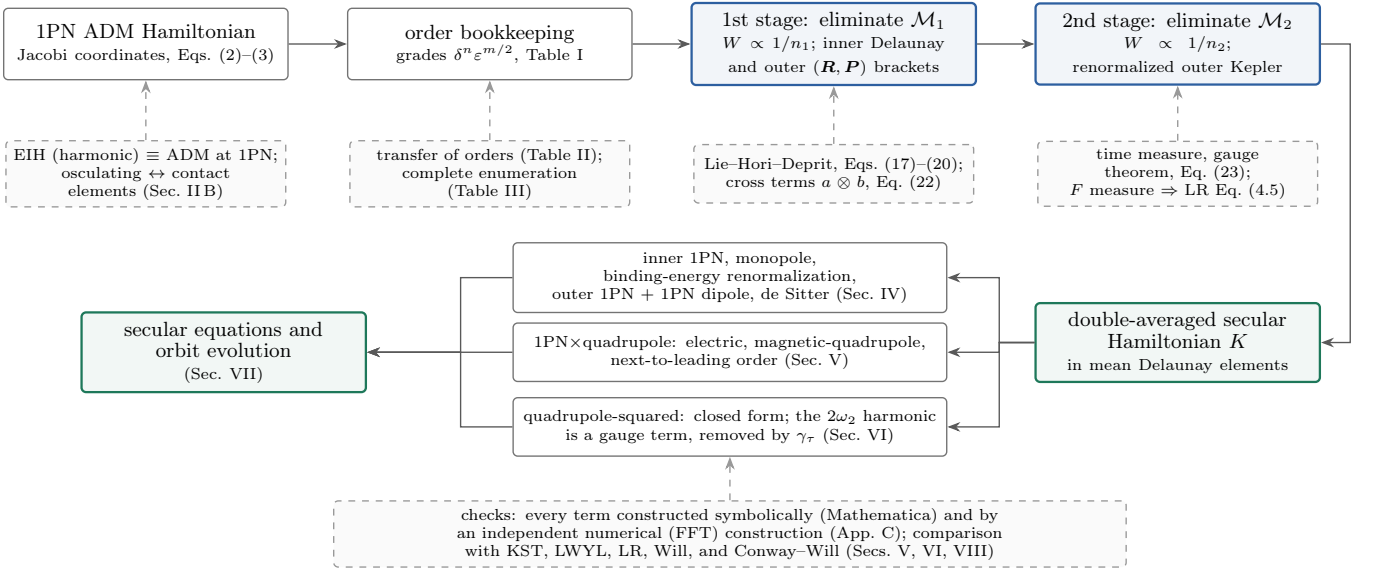

The structure of the calculation is summarized in Fig.~\ref{fig:flow}.

\subsection{Lie--Hori--Deprit transformation}

For functions $f,g$ of the canonical variables $(q_k,p_k)$ the Poisson bracket is
\begin{equation}
\{f,g\}=\sum_k\Big(\frac{\partial f}{\partial q_k}\frac{\partial g}{\partial p_k}
-\frac{\partial f}{\partial p_k}\frac{\partial g}{\partial q_k}\Big),
\label{eq:PB}
\end{equation}
so that $\dot f=\{f,H\}$. The sum runs over the canonical pairs in use: the Jacobi
pairs $(\bm r,\bm p)$ and $(\bm R,\bm P)$ in Cartesian form, or the Delaunay pairs
$(\M_i,L_i)$, $(\omega_i,G_i)$, $(\Omega_i,H_i)$ of Sec.~\ref{sec:ham-del}, with the
angles as the coordinates $q_k$ and the actions as the momenta $p_k$; in mixed form,
as in the first stage below, the inner pairs are Delaunay and the outer ones
Cartesian. With this sign convention the Lie operator of a generator $W$ is
\begin{equation}
\mathcal L_Wf=\{f,W\};
\label{eq:Lie}
\end{equation}
it is the Lie derivative along the Hamiltonian flow of $W$,
and the brackets that appear in the Lie--Hori--Deprit expansion below are all Poisson
brackets of the form of Eq.~\eqref{eq:PB}.

Let $H=H_0+\epsilon H_1+\epsilon^2H_2+\cdots$ and transform it with the Lie series
generated by $W=\epsilon W_1+\epsilon^2W_2+\cdots$ \cite{Hori1966,Deprit1969},
\begin{equation}
x^{\rm old}=\exp(\mathcal L_W)\,x^{\rm new}.
\label{eq:Lieseries}
\end{equation}
To second order the new Hamiltonian is
\begin{align}
K_1&=H_1+\{H_0,W_1\},\label{eq:K1}\\
K_2&=H_2+\tfrac12\{H_1+K_1,W_1\}+\{H_0,W_2\}.\label{eq:K2}
\end{align}
If $H_0$ depends only on the action $L$ conjugate to the angle $\M$ to be eliminated,
$\{H_0,W\}=-n\,\partial_{\M}W$ with $n=\partial H_0/\partial L$, and the homological
equation is solved by
\begin{gather}
n\,\partial_{\M}W_1=H_1-K_1\equiv\Delta_1,\qquad K_1=\av{H_1}_{\M},\nonumber\\
W_1=\frac1n\mathcal A_{\M}[\Delta_1]+\gamma,
\label{eq:homo}
\end{gather}
where $\mathcal A_{\M}$ is the mean-free antiderivative and $\gamma$ an arbitrary
function of the slow variables. With $\gamma=0$ the second-order secular Hamiltonian is
\begin{equation}
K_2=\av{H_2}_{\M}+\tfrac12\av{\{\Delta_1,W_1\}}_{\M},
\label{eq:K2avg}
\end{equation}
because
$\av{\{K_1,W_1\}}=-\partial_LK_1\av{\partial_{\M}W_1}+\{K_1,\av{W_1}\}_{\rm slow}$ vanishes.
Appendix~\ref{app:vz} compares this construction with the von Zeipel method used by
LWYL; to the orders needed here the two give the same $K$ in the same mean elements.

\paragraph*{Two small parameters.}
With the two parameters $\delta$ and $\varepsilon$ of Eq.~\eqref{eq:scal}, the perturbation
$H_1$ in Eq.~\eqref{eq:K1} is a sum of parts of different orders. Consider a 1PN part
$H^a\equiv H^{(1,a)}$ of order $(1,a)$ and a Newtonian part $H^b\equiv H^{(0,b)}$ of order
$(0,b)$ (Sec.~\ref{sec:ham-orders}), $H_1=H^a+H^b$; the superscripts $a$ and $b$ are order
labels, not powers. Each part $H^c$ ($c=a,b$) has its own oscillating part and
first-order generator,
\begin{equation}
\Delta^c=H^c-\av{H^c}_{\M},\qquad W^c=\frac1n\,\mathcal A_{\M}[\Delta^c],
\label{eq:DeltaW}
\end{equation}
so that $\Delta_1=\Delta^a+\Delta^b$ and $W_1=W^a+W^b$ in Eq.~\eqref{eq:homo}; in later sections
they are written with the full order label, e.g.\ $\Delta^{(1,0)}$ and $W^{(0,3)}$. The part of
$\tfrac12\av{\{\Delta_1,W_1\}}_{\M}$ in Eq.~\eqref{eq:K2avg} that couples the two is the
cross term
\begin{equation}
K^{a\otimes b}=\tfrac12\av{\{\Delta^a,W^b\}+\{\Delta^b,W^a\}}=\av{\{\Delta^a,W^b\}},
\label{eq:cross}
\end{equation}
since the bilinear form $B(a,b)=\av{\{\Delta^a,W^b\}}$ is symmetric
[integrate by parts in $\M$ and use $n\,\partial_{\M}W=\Delta$]. We use whichever
order is simpler; for the electric 1PN\,$\times$\,quadrupole term, for example,
$W$ of the quadrupole is a trigonometric polynomial in the eccentric anomaly, so that no
true anomaly is needed.

\paragraph*{Notation $a\otimes b$.}
The symbol $\otimes$ is not a tensor product. Throughout, $a\otimes b$ is shorthand
for the cross term, Eq.~\eqref{eq:cross}, between the 1PN part of order $(1,a)$, i.e.\ of
magnitude $\delta\varepsilon^{a}$, and the Newtonian part of order $(0,b)$, of magnitude
$\varepsilon^{b}$: one of the two enters through its oscillating part $\Delta$, the
other through its generator $W$, and the symmetry of $B(a,b)$ makes the choice
immaterial. Thus $0\otimes3$ couples the inner 1PN Hamiltonian $(1,0)$ to the
Newtonian quadrupole $(0,3)$, and $\tfrac12\otimes3$ the odd monopole term to the
quadrupole. In the first stage $\Delta$ and $W$ are those of $\M_1$ and the
Hamiltonians are the original ones; in the second stage they are those of $\M_2$ and
the Hamiltonians are the first-stage secular ones, $K^{(1,a)}$ and $K^{(0,b)}$. The
resulting term contributes to the order $(1,a+b+s)$, with the shift $s$ of
Table~\ref{tab:transfer} set by the variables in the bracket; for example, the
first-stage $0\otimes3$ is of order $(1,3)$, while the second-stage slow
$0\otimes3=\{K^{(1,0)},\av{W}\}$ is of order $(1,3/2)$. A subscript qualifies the
part of a term that is used, e.g.\ $2_{\rm odd}$ denotes the part of order
$(1,2)$ odd in $\bm r$ (the 1PN dipole; Sec.~\ref{sec:ham-eih}, item (iii), and Table~\ref{tab:names}).

\subsection{Gauge theorem and the averaging measure}
\label{sec:lie-gauge}

Replacing $W_1\to W_1+\gamma$ with $\gamma$ independent of $\M$ changes the
second-order Hamiltonian by a Poisson bracket with the first-order one,
\begin{equation}
K_2\ \to\ K_2+\{K_1,\gamma\},
\label{eq:gaugethm}
\end{equation}
as follows from $\av{\tfrac12\{H_1+K_1,\gamma\}}=\{K_1,\gamma\}$. The canonical
(Brouwer) normalization is $\av{W_1}_{\M}=0$ with the \emph{time} measure. Imposing
mean-freedom with another measure $d\mu$ is equivalent to
$\gamma^{\mu}=-\av{W_1}_{\mu}\neq0$. For LR's measure, uniform in the outer true
anomaly, the second-stage cross term of the inner 1PN precession with the quadrupole
becomes
\begin{equation}
K^{(1,3/2)}\big|_F=\{K^{(1,0)},\gamma^F\}\neq0,\qquad
K^{(1,3/2)}\big|_t=0,
\label{eq:K32gauge}
\end{equation}
which is the canonical form of the conclusion of Paper~I: LR's Eq.~(4.5) is a pure
gauge term, one term of the flow of $\{K^{(1,0)},\gamma^F\}$. The same theorem controls
the $2\omega_2$ harmonic of the quadrupole-squared term (Sec.~\ref{sec:q2-gauge}).

\subsection{Two stages and the transfer of orders}
\label{sec:lie-stages}

We eliminate first the inner mean anomaly $\M_1$ and then the outer one $\M_2$.
Dimensional analysis gives the transfer rules of Table~\ref{tab:transfer}: in the
first stage $W\propto1/n_1$ and the brackets with respect to the inner and slow variables
preserve the order, whereas brackets with respect to the outer Cartesian variables
raise it by $1/2$; in the second stage $W\propto1/n_2\propto a_2^{3/2}$, so that slow
brackets lower the order by $3/2$ and brackets with respect to $(\M_2,L_2)$, which
carry an extra $\partial_{L_2}\propto a_2^{-1/2}$, lower it by $1$.

\begin{table}[tb]
\caption{Transfer of orders in the two stages: a cross term of two terms of orders
$(1,a)$ and $(0,b)$ contributes to $(1,a+b+s)$ with the shift $s$ listed.}
\label{tab:transfer}
\centering\small
\begin{tabular}{@{}llc@{}}
\toprule
Stage & Bracket & $s$\\
\midrule
1st ($\M_1$ eliminated) & inner and slow variables & $0$\\
 & outer $(\bm R,\bm P)$ & $+\tfrac12$\\
2nd ($\M_2$ eliminated) & slow $(\omega_i,G_i)$ & $-\tfrac32$\\
 & fast $(\M_2,L_2)$ & $-1$\\
\bottomrule
\end{tabular}
\end{table}

\subsection{Complete enumeration}
\label{sec:lie-enum}

Since the Newtonian interaction starts at $(0,3)$, the rules leave only a few
contributions to each order. Table~\ref{tab:enum} lists them; ``$=0$'' marks those
that vanish identically, for the reasons given. The Hamiltonian of the second stage is
taken to be the \emph{renormalized} outer Kepler Hamiltonian of Sec.~\ref{sec:mono-11};
the fast second-stage cross terms of the binding-energy term are then absorbed into the
outer elements.

\begin{table*}[tb]
\caption{Complete enumeration of the contributions to each 1PN order. ``Direct'' is
the double average of the Hamiltonian of that order; $a\otimes b$ denotes a cross term
between the orders $(1,a)$ and $(0,b)$. The electric, magnetic and next-to-leading order
1PN\,$\times$\,quadrupole terms are the orders $(1,3)$, $(1,7/2)$ and $(1,4)$.}
\label{tab:enum}
\centering\small
\begin{tabular}{@{}lp{0.82\textwidth}@{}}
\toprule
Order & Contributions\\
\midrule
$(1,1)$ & direct only\\
$(1,\tfrac32)$ & direct $=0$ ($H^{(1,3/2)}\equiv0$); 2nd stage, slow: $0\otimes3=\{K^{(1,0)},\av{W}_t\}=0$\\
$(1,2)$ & direct; 2nd stage, fast: $0\otimes3=0$ ($K^{(1,0)}$ independent of $L_2$)\\
$(1,\tfrac52)$ & direct; 2nd stage, slow: $1\otimes3=0$ (structure theorem), $0\otimes4=0$ (time gauge);
 fast: $\tfrac12\otimes3=0$ ($K^{(1,1/2)}=0$)\\
$(1,3)$ & direct; \textbf{1st stage} $0\otimes3$; 2nd stage, slow: $\tfrac32\otimes3=0$; fast: $1\otimes3$ absorbed
 in the renormalized outer orbit, $0\otimes4=0$\\
$(1,\tfrac72)$ & direct; 1st stage $\tfrac12\otimes3=0$ (Sec.~\ref{sec:pnq-m}); 1st stage, outer: $0\otimes3=0$;
 \textbf{2nd stage, slow} $2_{\rm odd}\otimes3$; $1\otimes4=0$ (structure theorem), $0\otimes5=0$ (time gauge);
 fast: $\tfrac32\otimes3=0$\\
$(1,4)$ & direct; 1st stage $1\otimes3$, $0\otimes4$; 1st stage, outer $\tfrac12\otimes3$;
 2nd stage, slow: $\tfrac52\otimes3$, $\tfrac32\otimes4=0$; fast: $2\otimes3$, $1\otimes4$ absorbed\\
\bottomrule
\end{tabular}
\end{table*}

\subsection{Tools}

Three devices make the calculation tractable (Appendix~\ref{app:pipe}).
(i) The orientation is handled abstractly: with the inner basis
$(\hat{\bm\alpha},\hat{\bm\beta},\hat{\bm\gamma})$ (pericenter, in-plane normal, orbit
normal) and the projections $A_a,B_a,G_a$ of the outer basis vectors on it,
$\partial_{\omega_1}:A\to B,\ B\to-A,\ G\to0$ and
$\partial_\iota:A\to\sin\omega_1G,\ B\to\cos\omega_1G,\ G\to-(\sin\omega_1A+\cos\omega_1B)$.
(ii) The first-order generator of the outer elimination is written in closed form as
\begin{gather}
n_2W^*=\tilde G+c_0\,v,\nonumber\\
\tilde G=\int\bar H\,w_2\,dF-c_0F,\qquad v=F-\M_2,
\label{eq:Wstar}
\end{gather}
where $\bar H=\av{H^{(0,3)}}_{\M_1}$, $c_0=\av{\bar H}_{\M_2}$ and $\tilde G$ is a
trigonometric polynomial in $F$. (iii) Averages involving the equation of the center $v$
are reduced by the identity
\begin{equation}
\av{v\,h}_{\M_2}=\av{J}_{\M_2}-\av{J}_{F},\qquad J=\int h\,d\M_2,
\label{eq:vh}
\end{equation}
valid for $\av{h}_{\M_2}=0$, where $\av{\cdot}_F$ is the uniform average in $F$.

\section{Monopole and dipole terms}
\label{sec:mono}

\subsection{Inner 1PN precession}
\label{sec:mono-10}

The double average of the two-body part gives the standard result
\begin{gather}
K^{(1,0)}=-\frac{G^2m^2\mu_1}{c^2a_1^2}\left[\frac{3}{\ell_1}-\frac{15-\nu}{8}\right],
\nonumber\\
\dot\omega_1=\frac{\partial K^{(1,0)}}{\partial G_1}=\frac{3(Gm)^{3/2}}{c^2a_1^{5/2}\ell_1^2}.
\label{eq:K10}
\end{gather}

\subsection{The monopole term odd in the outer momentum}
\label{sec:mono-half}

The part of the Hamiltonian of order $\delta\varepsilon^{1/2}$ is linear in $\bm P$ and odd
in the inner variables. Only the $\bm P$-linear part of $-\frac18\sum p_a^4/m_a^3$ and
the inner pair terms contribute (terms with $r_{13},r_{23}$ are of order
$\tau^3$ or higher), and its inner average reduces to the three vector averages
$\av{p^2\bm p}$, $\av{\bm p/r}$ and $\av{(\hat{\bm n}\!\cdot\!\bm p)\hat{\bm n}/r}$.
The Kepler identities
\begin{gather}
\av{\frac{d\hat{\bm n}}{dt}}=\av{\frac{\bm v-\dot r\hat{\bm n}}{r}}=0,\nonumber\\
\av{v^2\bm v}=Gm\av{\Big(\frac2r-\frac1{a_1}\Big)\bm v}=2Gm\av{\frac{\bm v}{r}}
\end{gather}
make all three proportional to $\bm U\equiv\av{\bm v/r}\neq0$. The coefficient of
$\bm U\cdot\bm P$ is $G\mu_1(m_2-m_1)/m$ from the kinetic term and $G\mu_1(m_1-m_2)/m$ from
the pair terms, so that
\begin{equation}
\av{H^{(1,1/2)}}_{\M_1}=0
\label{eq:K12}
\end{equation}
identically. The oscillating part, however, generates a first-stage generator
$W^{(1,1/2)}$ that enters the magnetic-quadrupole and next-to-leading order
1PN\,$\times$\,quadrupole terms.

\subsection{Binding-energy renormalization}
\label{sec:mono-11}

The part of order $\delta\varepsilon$, averaged over the inner orbit, is
\begin{gather}
\av{H^{(1,1)}}_{\M_1}=\frac{|E_b|}{c^2}\left[\frac{P^2}{2m^2}+\frac{Gm_3}{R}\right],
\nonumber\\
E_b=-\frac{Gm_1m_2}{2a_1}\equiv-|E_b|.
\label{eq:h11}
\end{gather}
The $P^2$ coefficient collects the expansion of $-\frac18\sum p_a^4/m_a^3$ and the
inner pair term; the tensor virial theorem $\av{\mu_1v^iv^j}=\av{Gm\mu_1n^in^j/r}$%
\footnote{For a bound Kepler orbit the function $\mu_1r^iv^j$ is periodic, so the time
average of its derivative vanishes. With $\mu_1\dot v^j=-Gm\mu_1n^j/r^2$ this gives
$\av{\mu_1v^iv^j}=\av{Gm\mu_1r^in^j/r^2}=\av{Gm\mu_1n^in^j/r}$, the tensor form of the
virial theorem; its trace is the familiar $2\av{T}=-\av{V}$. Explicitly,
$\av{v^iv^j}=Gm\big(\hat\alpha^i\hat\alpha^j+\ell_1\hat\beta^i\hat\beta^j\big)/[a_1(1+\ell_1)]$.}
removes the anisotropic parts. The $1/R$ coefficient is
$-\frac{3G}{2}m_3\av{p^2}(m_1^{-1}+m_2^{-1})+2G^2m_1m_2m_3\av{1/r}=G^2m_1m_2m_3/(2a_1)$.
Eq.~\eqref{eq:h11} states that
\begin{equation}
\frac{P^2}{2\mu_2}-\frac{Gmm_3}{R}+\av{H^{(1,1)}}_{\M_1}
=\frac{P^2}{2\mu_2'}-\frac{Gm'm_3}{R}+\mathcal{O}(c^{-4}),
\label{eq:renorm}
\end{equation}
with $m'=m-|E_b|/c^2$ in both the inertial and the gravitational mass: the outer Kepler
problem sees the inner binary with its rest mass reduced by the binding energy. It
follows that the term depends neither on the orientation nor on $e_1$, that its
second-stage slow cross terms vanish identically (the ``structure theorem'' used in
Table~\ref{tab:enum}), and that its fast cross terms are absorbed if the second stage
is built on the renormalized outer Kepler Hamiltonian. Averaged over the outer orbit,
\begin{equation}
K^{(1,1)}=\frac{G^2m_1m_2m_3\,(2m+3m_3)}{4c^2M\,a_1a_2},
\end{equation}
a function of $L_1$ and $L_2$ that shifts the mean motions only.

\paragraph*{Equivalence principle.}
Repeating the calculation with the fully conservative PPN $N$-body Lagrangian (the EIH
form with parameters $\beta,\gamma$ \cite{Will2018book}) gives, to twelve digits in
numerical checks for several $(\beta,\gamma)$,
\begin{gather}
\av{H^{(1,1)}}_{\M_1}=-\frac{E_b}{c^2}\frac{P^2}{2m^2}
-\frac{Gm_3}{R}\frac{E_b}{c^2}\,(1+2\eta_N),\nonumber\\
\eta_N=4\beta-\gamma-3,
\label{eq:h11ppn}
\end{gather}
i.e.\ the inertial mass $m+E_b/c^2$ is independent of $\beta,\gamma$, and the
gravitational mass is $m+E_b/c^2+\eta_N\Omega_b/c^2$ with the gravitational binding energy
$\Omega_b=2E_b$. This is Nordtvedt's result \cite{Nordtvedt1968} for a gravitationally
bound composite: the binding-energy term probes the \emph{strong} equivalence principle,
and in GR it expresses the equality of the inertial and gravitational mass of the
binding energy. As a static effect it is degenerate with the masses, and it is far
too small to compete with lunar laser ranging \cite{HofmannMullerBiskupek2010}; we
mention it because it explains the structure of the cross terms. Note also that the
individual-body analogue, a difference of $m_{\rm G}/m_{\rm I}$ between $m_1$ and $m_2$,
would produce a dipole term with the same angular structure as the 1PN dipole of
Sec.~\ref{sec:mono-2}; for the triple pulsar PSR~J0337+1715 the GR dipole mimics a
violation parameter $\Delta_{\rm eff}=G(m_1-m_2)/(6a_1c^2)\simeq6\times10^{-8}$, well
below the current bound \cite{Voisin2020}.

\subsection{The order $\delta\varepsilon^{3/2}$}
\label{sec:mono-32}

The $\tau^3$ part of Eq.~\eqref{eq:adm} consists of kinetic terms of the type
$(\bm p\!\cdot\!\bm P)P^2$ and pair terms of the type $(\bm p\!\cdot\!\bm P)/R$. The former
cancel between bodies~1 and~2,
$-\frac18[-4(m_1/m)^3/m_1^3+4(m_2/m)^3/m_2^3](\bm p\!\cdot\!\bm P)P^2=0$, and the latter
cancel because $\bm p_1\simeq\bm p$, $\bm p_2\simeq-\bm p$ and $1/r_{13}=1/r_{23}+\mathcal{O}(\varepsilon)$.
Hence $H^{(1,3/2)}\equiv0$ before any average (confirmed by the full symbolic $\tau$
expansion), and by Eq.~\eqref{eq:K32gauge} the only candidate cross term vanishes in
the time gauge:
\begin{equation}
K^{(1,3/2)}=0 .
\end{equation}
The libration term of LR is absent from the canonical secular Hamiltonian at every
stage, consistent with KST and LWYL and with Paper~I.

\subsection{Outer 1PN precession and the 1PN dipole}
\label{sec:mono-2}

The inner average of the part of order $\delta\varepsilon^2$ is
\begin{equation}
\begin{split}
\av{H^{(1,2)}}_{\M_1}={}&H^{\rm 1PN}_{\rm 2b}(m,m_3;\bm R,\bm P)\\
&+\frac{G^2m_1m_2(m_1-m_2)m_3\,e_1}{4mc^2}\,\frac{\hat{\bm\alpha}\!\cdot\!\bm R}{R^3},
\end{split}
\label{eq:h12}
\end{equation}
where the first term is exactly the ADM two-body 1PN Hamiltonian of masses $m$ and
$m_3$. Its double average gives the outer precession,
\begin{equation}
K^{(1,2)}=-\frac{G^2M^2\mu_2}{c^2a_2^2}\left[\frac{3}{\ell_2}-\frac{15-\mu_2/M}{8}\right].
\label{eq:K1two}
\end{equation}
The second term is the
inner average of the 1PN dipole, proportional to $\av{\bm r}=-\frac32a_1e_1\hat{\bm\alpha}$
and to the octupole mass factor; it averages to zero over the outer orbit
($\av{\bm R/R^3}_{\M_2}=0$), but its oscillating part is responsible for the
second-stage part of the magnetic-quadrupole term. It is the canonical counterpart of
the 1PN correction to the center of mass of the inner binary.

\subsection{The de Sitter term}
\label{sec:mono-dS}

The only contribution of order $\delta\varepsilon^{5/2}$ is direct. Its inner average is
the spin--orbit coupling of the inner orbital angular momentum $\bm G_1=\bm r\times\bm p$
with the outer orbit,
\begin{equation}
\av{H^{(1,5/2)}}_{\M_1}=\frac{G(4m+3m_3)}{2mc^2R^3}\,(\bm R\times\bm P)\cdot\bm G_1,
\end{equation}
and its double average is
\begin{equation}
K^{(1,5/2)}=\frac{G(4m+3m_3)}{2mc^2}\,\frac{G_1G_2\cos\iota}{a_2^3\ell_2^3}.
\label{eq:KdS}
\end{equation}
Since $G_2\propto\mu_2=mm_3/M$, the prefactor is proportional to $m_3(4m+3m_3)$, as in
LR Eq.~(4.4) and KST; it vanishes for $m_3\to0$.

\section{1PN\,$\times$\,quadrupole cross terms}
\label{sec:pnq}

The three 1PN\,$\times$\,quadrupole terms correspond to the three velocity factors
$v^2$, $vV$ and $V^2$ that multiply the quadrupole tidal field. We write the ZLK
Hamiltonian as
\begin{equation}
K_{\rm Q}=A_{\rm Q}\,Z,\qquad
A_{\rm Q}=-\frac{G\mu_1m_3a_1^2}{16a_2^3\ell_2^3},
\label{eq:KQ}
\end{equation}
with the ZLK angular structure
\begin{equation}
\begin{split}
Z&=(2+3e_1^2)(3\cos^2\iota-1)+15e_1^2\sin^2\iota\cos2\omega_1\\
&=-2+12e_1^2+6(\bm j\!\cdot\!\hat{\bm z}_2)^2-30(\bm e\!\cdot\!\hat{\bm z}_2)^2,
\end{split}
\label{eq:Z}
\end{equation}
where $\bm j=\ell_1\hat{\bm\gamma}$ and $\bm e=e_1\hat{\bm\alpha}$ are the inner
angular-momentum and eccentricity vectors and $\hat{\bm z}_2$ is the outer orbit normal.

\subsection{Electric 1PN\,$\times$\,quadrupole term}
\label{sec:pnq-e}

By Table~\ref{tab:enum} the order $\delta\varepsilon^3$ receives the direct term and the
first-stage cross term of the inner 1PN Hamiltonian with the quadrupole,
\begin{equation}
K^{(1,3)}=\av{H^{(1,3)}}_{\M_1,\M_2}+\av{\{\Delta^{(1,0)},W^{(0,3)}\}}_{\M_1,\M_2},
\end{equation}
while the fast second-stage cross term of the binding-energy term is absorbed in the
renormalized outer orbit (Sec.~\ref{sec:mono-11}). The direct term is independent of
$\bm P$; its inner average, in the components $(X,Y,Z)$ of $\bm R$ along
$(\hat{\bm\alpha},\hat{\bm\beta},\hat{\bm\gamma})$, is given in Appendix~\ref{app:expr},
together with that of the cross term. For the cross term we use the symmetry
of Eq.~\eqref{eq:cross} and bracket $\Delta^{(1,0)}$ with $W^{(0,3)}$, which is a trigonometric
polynomial in the eccentric anomaly; since $\Delta^{(1,0)}$ does not depend on the
orientation, only the pairs $(\M_1,L_1)$ and $(\omega_1,G_1)$ contribute. With
$\av{R^iR^j/R^5}_{\M_2}=(\delta^{ij}-\hat z_2^i\hat z_2^j)/(2a_2^3\ell_2^3)$ the double
average is
\begin{equation}
\begin{split}
K^{(1,3)}={}&\frac{G^2m_1m_2m_3\,a_1}{c^2a_2^3\ell_2^3}
\Big[F_0(e_1)(1-3\cos^2\iota)\\
&+F_2(e_1)\sin^2\iota\cos2\omega_1\Big],
\end{split}
\label{eq:K13}
\end{equation}
\begin{align}
F_0&=\frac{1}{32}\Big[54+3e_1^2-\frac{84+36e_1^2}{\ell_1}-\nu(10-13e_1^2)\Big],
\label{eq:F0}\\
F_2&=\frac{3}{32}\Big[3e_1^2(1-\nu)\nonumber\\
&\quad-\frac{2(2+3e_1^2)\big[2(\ell_1-1)+e_1^2(4\ell_1-3)\big]}{e_1^2\ell_1}\Big].
\label{eq:F2}
\end{align}
The split of $F_0$ and $F_2$ between the direct term and the first-stage cross term
$0\otimes3$ is given in Table~\ref{tab:K13split}.
For $e_1\to0$, $F_0\to-(30+10\nu)/32$ and $F_2\to9e_1^2(4-\nu)/32$. Like the ZLK term,
$K^{(1,3)}$ is independent of $\omega_2$, and it vanishes for $m_3\to0$.

\begin{table}[tb]
\caption{Split of the coefficients $F_0$ and $F_2$ of the electric 1PN\,$\times$\,quadrupole term, Eqs.~\eqref{eq:F0}--\eqref{eq:F2}, into the direct term (double average of the 1PN Hamiltonian at quadrupole order) and the first-stage cross term $0\otimes3$ of the inner 1PN Hamiltonian with the Newtonian quadrupole.}
\label{tab:K13split}
\centering\small
\renewcommand{\arraystretch}{1.25}
\begin{tabular}{@{}lll@{}}
\toprule
 & Part & Expression\\
\midrule
$F_0$ & direct & $\dfrac{1}{32}\big[2-5e_1^2-\nu(10-13e_1^2)\big]$\\[2mm]
 & cross & $\dfrac18\Big[13+2e_1^2-\dfrac{21+9e_1^2}{\ell_1}\Big]$\\[2mm]
$F_2$ & direct & $\dfrac{9}{32}\,e_1^2(1-\nu)$\\[2mm]
 & cross & $-\dfrac{3(2+3e_1^2)\big[2(\ell_1-1)+e_1^2(4\ell_1-3)\big]}{16e_1^2\ell_1}$\\
\bottomrule
\end{tabular}
\end{table}

\paragraph*{Comparison with KST.}
The first term of KST Eq.~(35) contains the 1PN correction of the electric quadrupole,
$f_{\alpha,\beta}=\cdots-(Gm/c^2a_1)\,g_{\alpha,\beta}(e_1,\nu)$ [their
Eqs.~(23)--(25)]. In our normalization it gives $F_0^{\rm KST}=-(g_\alpha+g_\beta)/16$ and
$F_2^{\rm KST}=3(g_\alpha-g_\beta)/16$, which coincide with
Eqs.~\eqref{eq:F0}--\eqref{eq:F2} identically in $e_1$ and $\nu$. KST work in contact
elements with a time-mean-free normalization; at this order these coincide with our
canonical mean elements, so that the agreement is exact.

\subsection{Magnetic-quadrupole term}
\label{sec:pnq-m}

The order $\delta\varepsilon^{7/2}$ has two nonvanishing contributions
(Table~\ref{tab:enum}): the direct term, which comes from the $\bm p_a\cdot\bm p_b$ terms of
Eq.~\eqref{eq:adm} expanded to quadrupole order and from the $G^2$ three-body terms,
and the second-stage slow cross term of the odd (dipole) part of the order
$\delta\varepsilon^2$ with the quadrupole. The result is
\begin{equation}
\begin{split}
K^{(1,7/2)}={}&\underbrace{\frac{4m+3m_3}{4M}\Kmq}_{\rm direct}
+\underbrace{\frac{m_3}{4M}\Kmq}_{2_{\rm odd}\otimes3}\\
&+\underbrace{0}_{\frac12\otimes3}+\underbrace{0}_{1\otimes4}=\Kmq,
\end{split}
\label{eq:K72}
\end{equation}
with
\begin{equation}
\Kmq=-\frac{3G^2m_1m_2(m_1-m_2)\,m_3M^{1/2}\,e_1e_2\,\ell_1\,a_1^{3/2}}
{c^2m^{3/2}a_2^{7/2}\ell_2^4}\,\mathcal B,
\label{eq:Kmq}
\end{equation}
\begin{equation}
\mathcal B=\cos\iota\cos\omega_1\cos\omega_2+(2\cos^2\iota-1)\sin\omega_1\sin\omega_2 .
\label{eq:B}
\end{equation}
In vector form, with $\bm e_1$ the eccentricity vector of $\bm r=\bm x_1-\bm x_2$, $\bm e_2$ that
of $\bm R$, and $\bm j_i=\ell_i\hat{\bm z}_i$,
\begin{equation*}
\begin{split}
\Kmq={}&-\frac{3G^2m_1m_2(m_1-m_2)\,m_3M^{1/2}a_1^{3/2}}{c^2m^{3/2}a_2^{7/2}\ell_2^5}\\
&\times\big[(\bm j_1\times\bm e_1)\cdot(\bm j_2\times\bm e_2)
-2(\bm j_1\cdot\bm j_2)(\bm e_1\cdot\bm e_2)\big],
\end{split}
\end{equation*}
which does not depend on the choice of angles; the overall sign of Eq.~\eqref{eq:Kmq} refers to
the angles of Sec.~\ref{sec:ham-del}. $\Kmq$ is the magnetic-quadrupole term of KST Eq.~(35),
including the sign and the mass factor $m_3M^{1/2}(X_1-X_2)$, $X_i=m_i/m$, once it is noted
that KST's $\bm R$ points from the tertiary to the inner binary (so that their $\bm\alpha_3=-\hat{\bm e}_2$)
and that their averaged Lagrangian enters the Hamiltonian with the opposite sign. (The rate
$\dot p_1=2p_1\dot G_1/G_1$ quoted as Eq.~(A-p1KST) in Paper~I has the opposite overall sign,
$X_2-X_1$ in place of $X_1-X_2$; the implementation of Paper~I uses the correct sign.)
The canonical derivation thus confirms the KST result independently.

The two vanishing cross terms deserve comment. The first-stage cross term
$\Delta_1\equiv\av{\{H^{(0,3)},W^{(1,1/2)}\}}$ is not zero after the inner average (it is
proportional to $e_1(m_1-m_2)$), but vanishes after the outer average. To see this, one
must note that $\partial/\partial\Omega_1=-\partial/\partial\omega_2$ acts also on the outer
factors, so that three outer tensors are needed,
\begin{gather}
\Lambda^{kij}=\av{P^k\mathcal T^{ij}},\nonumber\\
\Lambda_A^{kij}=\av{(\partial_{\omega_2}P^k)\mathcal T^{ij}},\quad
\Lambda_B^{kij}=\av{P^k\partial_{\omega_2}\mathcal T^{ij}},
\end{gather}
with $\mathcal T^{ij}=\hat N^i\hat N^j/R^3$. With the abstract orientation algebra the
inner factors combine with these tensors to give zero identically. (Keeping only
$\Lambda$, as one might after noticing that $\Lambda^{000}=0$, would be wrong:
$\Lambda_A^{000}=7/8$.) The second-stage cross term of the binding-energy term with the
octupole vanishes before averaging, because Eq.~\eqref{eq:h11} depends neither on the
orientation nor on $e_1$.

Several independent checks support Eq.~\eqref{eq:K72}. The zero cross terms were
confirmed at eight independent configurations with 50-digit arithmetic. The split
direct\,:\,cross $=(4m+3m_3):m_3$ was reproduced by the numerical construction of
Sec.~\ref{sec:pnq-n}, in which the second-stage cross term of order $a_2^{-7/2}$ agrees with
$\Kmq\,m_3/(4M)$ in angular structure, prefactor and sign. And evaluating the direct term with
the nine combinations of time, eccentric-anomaly and true-anomaly measures for the two orbits
reproduces KST only for the time measures, and never approaches LR's (A7)--(A11) (Paper~I).
The magnetic-quadrupole term is the canonical form of the 1PN center-of-mass correction:
the part $m_3/(4M)$ arises from the dipole of Sec.~\ref{sec:mono-2}, which exists only because
$\bm R$ is referred to the Newtonian center of mass.

\subsection{Next-to-leading order 1PN\,$\times$\,quadrupole term}
\label{sec:pnq-n}

The order $\delta\varepsilon^4$ has the largest number of contributions
(Table~\ref{tab:enum}): the direct term; the first-stage cross terms of the
binding-energy term with the quadrupole and of the inner 1PN Hamiltonian with the
octupole; the first-stage cross term of the odd monopole term with the quadrupole through
the outer Cartesian bracket; and the second-stage cross terms of the de Sitter term (slow)
and of the outer 1PN Hamiltonian and the 1PN dipole (fast) with the quadrupole.

\paragraph*{First stage.}
{\sloppy We add to the inner Delaunay bracket the outer Cartesian bracket
\begin{equation}
\{F,W\}_{\rm out}=\partial_{\bm R}F\cdot\partial_{\bm P}W
-\partial_{\bm P}F\cdot\partial_{\bm R}W
\label{eq:PBout}
\end{equation}
and let the $\tau$ scaling (with a factor $\tau$ for the outer bracket) assign the orders
automatically. The first-stage cross term of the binding-energy term with the quadrupole is
not zero: it is built from the \emph{oscillating} part of $H^{(1,1)}$ (the instantaneous
$p^2$ and $1/r$), not from its average.\par}

\paragraph*{Second stage.}
The second stage is carried out with the exact pairs $(\omega_1,G_1)$, $(\omega_2,G_2)$ and
$(\M_2,L_2)$ at fixed $J$. The terms $1/G_2$ of Eq.~\eqref{eq:dcosi}, which are dropped
in the test-particle frame, are of relative order $\varepsilon^{1/2}$ and contribute to
this order. The generator $W^*$ of Eq.~\eqref{eq:Wstar} is a trigonometric polynomial in
$F$ plus a term linear in $v$; the latter is averaged with Eq.~\eqref{eq:vh}. The
orientation factors are kept separate from the functions of the anomalies throughout.

\ifreprint\widetext\fi 
\paragraph*{Result.}
With $U_4\equiv G^2m_1m_2m_3a_1^2/(c^2a_2^4\ell_2^5)$ the part symmetric in
$m_1\leftrightarrow m_2$ (``quadrupole type'') is
\begin{equation}
K^{(1,4)}_{\rm quad}=U_4\big[c_Z\,Z+c_Y\,Y\big],
\label{eq:K14}
\end{equation}
where $Z$ is the ZLK structure of Eq.~\eqref{eq:Z} and
\begin{equation}
\begin{split}
Y={}&(2+3e_1^2)\sin^2\iota\cos2\omega_2
+10e_1^2\Big[\cos^4\tfrac{\iota}{2}\cos(2\omega_1-2\omega_2)
+\sin^4\tfrac{\iota}{2}\cos(2\omega_1+2\omega_2)\Big]\\
={}&10\big[(\bm e\!\cdot\!\hat{\bm u}_2)^2-(\bm e\!\cdot\!\hat{\bm v}_2)^2\big]
-2\big[(\bm j\!\cdot\!\hat{\bm u}_2)^2-(\bm j\!\cdot\!\hat{\bm v}_2)^2\big],
\end{split}
\label{eq:Y}
\end{equation}
with $\hat{\bm u}_2$ the outer pericenter direction and
$\hat{\bm v}_2=\hat{\bm z}_2\times\hat{\bm u}_2$. The coefficients are
\begin{align}
c_Z&=-\frac{(7+18\ell_2+29e_2^2)\,m^2+(11+36\ell_2+52e_2^2)\,mm_3+(3+18\ell_2+27e_2^2)\,m_3^2}{32\,mM},
\label{eq:cZ}\\
c_Y&=-\frac{1-\ell_2}{1+\ell_2}\,
\frac{12(3\ell_2^2+2\ell_2+1)\,m^2+2(39\ell_2^2+46\ell_2+23)\,mm_3+6(7\ell_2^2+10\ell_2+5)\,m_3^2}{128\,mM}.
\label{eq:cY}
\end{align}
\ifreprint\endwidetext\fi
The contributions of the individual terms are listed in Appendix~\ref{app:expr}. The first
term renormalizes the ZLK coupling. The second is proportional to $e_2^2$
[$(1-\ell_2)/(1+\ell_2)\simeq e_2^2/4$] and depends on $\omega_2$, because the outer average
$\av{P^iP^jR^kR^l/R^5}$ is anisotropic in the outer plane when $e_2\neq0$; the Newtonian
quadrupole has no such structure. Consequently the component of the inner angular momentum
normal to the outer orbit, which the ZLK Hamiltonian conserves, is not conserved at this
order. The de Sitter\,$\times$\,ZLK cross term contributes only to $c_Y$.

The part antisymmetric in $m_1\leftrightarrow m_2$ (``octupole type'') is
$-U_4\,e_2(m_1-m_2)m^{-3}[c_3T_3+c_1T_1]$, with the angular structures
\begin{align}
T_3&=\sin^2\iota\Big[\cos^2\tfrac\iota2\cos(3\omega_1-\omega_2)
+\sin^2\tfrac\iota2\cos(3\omega_1+\omega_2)\Big],\nonumber\\
T_1&=\cos^2\tfrac\iota2\,(13-20\cos\iota+15\cos2\iota)\cos(\omega_1-\omega_2)\nonumber\\
&\quad+\sin^2\tfrac\iota2\,(13+20\cos\iota+15\cos2\iota)\cos(\omega_1+\omega_2),
\end{align}
where $c_3=\mathcal{O}(e_1^3)$ and $c_1=\mathcal{O}(e_1)$ are quadratic forms in $(m_1,m_2)$ with coefficients depending on $e_1$, regular at $e_1=0$ and independent of $m_3$ (Appendix~\ref{app:expr}). It
comes from the first-stage cross terms with the octupole and through the outer bracket,
from the direct term, and from the second-stage cross term of the 1PN dipole, which is
exactly twice the outer-bracket contribution. All terms are proportional to $m_3$ and
vanish for a massless tertiary.

\paragraph*{Independent verification.}
The symbolic result was checked by a numerical construction that uses none of the
Mathematica expressions (Appendix~\ref{app:pipe}): the ADM Hamiltonian is evaluated as an
analytic function of a complex $\tau$ [with $1/r_{a3}=\tau^2/\sqrt{(\bm R-\tau^2\bm x_a)^2}$]
and the grades are extracted by FFT on $|\tau|=\mathrm{const}$; the inner antiderivative
is taken by FFT and the brackets by central differences; the second stage uses the exact
reduced pairs with 128 points in $\M_2$. At two generic configurations the first-stage
cross terms agree to $10^{-8}$--$10^{-9}$, the direct terms to $10^{-13}$, and the
second-stage cross terms to $2\times10^{-9}$. In the course of this check we found and
corrected a sign error in the de Sitter\,$\times$\,ZLK contribution to $c_Y$ in our first
symbolic assembly (a branch of $\sqrt{e_2^2-1}$); the coefficients above are the corrected
ones, and the assembled $c_Z,c_Y,c_3,c_1$ reproduce the full grade to $10^{-16}$.

\paragraph*{Comparison with LR.}
LR's Eq.~(4.6), the leading term for $m\ll M$, gives
$\dot e_1\propto e_2^2\ell_2^2\,S_{\rm LR}-g_{e_2}\sin^2\iota\sin2\omega_1$ with
$S_{\rm LR}=(3+\cos2\iota)\cos2\omega_2\sin2\omega_1-4\cos\iota\cos2\omega_1\sin2\omega_2$ and
$g_{e_2}=6(8+3e_2^2+4e_2^4)$. The rate derived from Eq.~\eqref{eq:K14} contains the same two
structures ($\partial_{\omega_1}Y$ is $S_{\rm LR}/4$ up to a factor), but the coefficients
differ. The $\omega_2$-dependent coefficient of LR equals, at six random configurations to
$10^{-5}$, one half of the sum of our fast second-stage cross term (outer
1PN\,$\times$\,ZLK) and the gauge term $\{K^{(1,2)},\gamma^F\}$ that arises from the
true-anomaly normalization of the quadrupole periodic solution. This is suggestive of the
same kind of measure mismatch as in Paper~I, but we have not traced it in LR's
implementation. LR's Eq.~(4.6) does not contain the de Sitter\,$\times$\,ZLK cross term,
which dominates our $c_Y$ for $m\ll M$. The $\omega_2$-independent coefficient has
the opposite sign in LR ($-6(8+3e_2^2+4e_2^4)/\ell_2^2$ against our
$+2(3+18\ell_2+27e_2^2)$ in the same normalization); this part has the angular structure of
the ZLK term and changes under a redefinition of the outer elements at 1PN order and under
the osculating--contact map of Sec.~\ref{sec:ham-eih}, so its comparison requires the
elements to be converted first. The consequences for the orbital evolution are examined in
Sec.~\ref{sec:orbit}.

\section{Quadrupole-squared term}
\label{sec:q2}

\subsection{Secular Hamiltonian in closed form}
\label{sec:q2-closed}

The quadrupole-squared term has no Hamiltonian of its own ($H_2=0$) and comes from the
second-stage cross term of the quadrupole with itself; by Eq.~\eqref{eq:K2avg}
\begin{equation}
K^{(0,9/2)}=\tfrac12\av{\{\bar H,W^*\}}_{\M_2},\qquad \bar H=\av{H^{(0,3)}}_{\M_1},
\label{eq:K92}
\end{equation}
with $W^*$ of Eq.~\eqref{eq:Wstar} and the bracket in the slow variables. Because $\bar H$
contains only the harmonics $0$ and $1$ of $\omega_2$, the result has the structure
$K^{(0,9/2)}=P(e_1,\iota,\omega_1)+A\cos2\omega_2+B\sin2\omega_2$. Its
$\omega_2$-independent part $P$ agrees pointwise, to seven significant digits at four
independent configurations and for all $e_2$, with the Brown Hamiltonian in the form
\cite{Tremaine2023,LeiGrishin2025a}
\begin{equation}
\begin{split}
&K_{\rm Brown}=C\varepsilon_{21}F_{21},\qquad C=6A_{\rm Q},\\
&F_{21}=\tfrac38j_z\big(1-j_z^2+24e_1^2-15e_z^2\big),\\
&\varepsilon_{21}=\frac{n_2}{n_1}\,\frac{m_3}{M}\,\frac{1+\frac23e_2^2}{\ell_2^3},
\end{split}
\label{eq:brown}
\end{equation}
with $j_z=\ell_1\cos\iota$ and $e_z=e_1\sin\iota\sin\omega_1$.

To identify the $\omega_2$-dependent part we evaluated $K^{(0,9/2)}$ numerically on a
$6\times6$ grid in $(\omega_1,\omega_2)$ and projected on the Fourier harmonics. Only
$1$, $\cos2\omega_1$, $\cos2\omega_2$ and $\cos(2\omega_1\mp2\omega_2)$ appear (sine
components and higher harmonics are below $10^{-15}$), and at 28 configurations (seven
inclinations, two $e_1$, two $e_2$) and three sets of masses the coefficients follow the
closed form
\begin{equation}
K^{(0,9/2)}=C\varepsilon_{21}\Big[F_{21}+\frac{15}{32}\,
\frac{e_2^2\,\Hf(e_2)}{3+2e_2^2}\,\mathcal Y_{92}\Big],
\label{eq:K92closed}
\end{equation}
\begin{equation}
\begin{split}
\mathcal Y_{92}={}&\ell_1\Big[5e_1^2(1+\cos\iota)^2(3\cos\iota-2)\cos(2\omega_1-2\omega_2)\\
&+5e_1^2(1-\cos\iota)^2(3\cos\iota+2)\cos(2\omega_1+2\omega_2)\\
&-2(2-17e_1^2)\cos\iota\sin^2\iota\,\cos2\omega_2\Big],
\end{split}
\label{eq:Y92}
\end{equation}
\begin{equation}
\Hf(e_2)=1-\frac{2(1-e_2^2)}{5(1+\ell_2)^2}
\qquad[\Hf(0)=0.9,\ \Hf(1)=1],
\label{eq:Hfun}
\end{equation}
to a relative accuracy of $10^{-8}$ ($10^{-10}$ for the mass dependence); at a random
point the closed form and the numerical value differ by $8\times10^{-11}$.

\paragraph*{Relation to Will and to Conway and Will.}
The secular equations $\dot e_1=(\ell_1/e_1L_1)\,\partial K/\partial\omega_1$ and
$\dot\Omega_1=\partial K/\partial H_1$ derived from Eq.~\eqref{eq:K92closed} reproduce
the quadrupole-squared equations of Will \cite{Will2021} and of Conway and Will
\cite{ConwayWill2024} (their inner-element feedback terms): the angular factors
$(1\mp\cos\iota)^2(3\cos\iota\pm2)\sin(2\omega_1\pm2\omega_2)$ in $\dot e_1$,
$(1\pm\cos\iota)(1\mp9\cos\iota)$ and $2(2-17e_1^2)(1-3\cos^2\iota)$ in $\dot\Omega_1$,
the function $\Hf(e_2)$, and the relative weight
$3(3+2e_2^2):\tfrac52e_2^2\Hf(e_2)$ of the Brown and $2\omega_2$ terms in $\dot e_1$, which
we find numerically as $6.000000$ in the normalization of Eq.~\eqref{eq:K92closed}. Thus
Eq.~\eqref{eq:K92closed} is the Hamiltonian that generates Will's secular equations. The
function $\Hf$ is also related to Tremaine's coefficient $C_1(e_2)$ of the
$\omega_2$-dependent part for the mean anomaly as fictitious time [his Eq.~(61)]: using
$4\ell^3-4+6e^2+3e^4=(1-\ell)^2(3\ell^2+10\ell+5)$,
\begin{equation}
C_1(e_2)\,\ell_2^6=\frac{(1-\ell_2)(3\ell_2^2+10\ell_2+5)}{12(1+\ell_2)}
=\frac{5}{12}\,e_2^2\,\Hf(e_2).
\label{eq:C1H}
\end{equation}

\paragraph*{Consequence for $j_z$.}
With the Brown term alone, $j_z=H_1/L_1$ is conserved. The $2\omega_2$ harmonic gives, in the
test-particle frame with $\omega_2=-\Omega_1$ up to a constant,
\begin{equation}
\dot H_1=\frac{\partial K}{\partial\omega_2}=2\,\mathrm{Im}\big[\Psi e^{2i\Omega_1}\big],
\qquad \Psi\equiv A+iB .
\label{eq:H1dot}
\end{equation}
Whether this is physical or a property of the mean elements is the question of the next
subsection.

\subsection{Is the $2\omega_2$ harmonic a gauge term?}
\label{sec:q2-gauge}

By Eq.~\eqref{eq:gaugethm} the harmonic can be removed by a change of mean elements if
there is a function $c$ of the slow variables with
\begin{equation}
A\cos2\omega_2+B\sin2\omega_2=\{K_{\rm Q},c\}.
\label{eq:gaugeeq}
\end{equation}
Since $K_{\rm Q}$ depends neither on $\Omega_1$ nor on $\omega_2$,
\begin{equation}
\begin{split}
\{K_{\rm Q},c\}&=\mathcal Dc+\dot\Omega_1\partial_{\omega_2}c,\\
\mathcal D&=\partial_{\omega_1}K_{\rm Q}\,\partial_{G_1}-\partial_{G_1}K_{\rm Q}\,\partial_{\omega_1}.
\end{split}
\label{eq:Dop}
\end{equation}
With $c=p\cos2\omega_2+q\sin2\omega_2$ and $\zeta=p+iq$ this becomes
\begin{equation}
\mathcal D\zeta-2i\dot\Omega_1\zeta=\Psi,
\label{eq:zetapde}
\end{equation}
two coupled linear first-order partial differential equations. Since $\mathcal D=-d/dt$ along the ZLK flow, the characteristics are
the ZLK orbits, along which
\begin{equation}
\frac{d}{dt}\big(e^{2i\Omega_1}\zeta\big)=-e^{2i\Omega_1}\Psi .
\label{eq:zetaode}
\end{equation}
Integrating over one ZLK period $T$,
\begin{equation}
\big(e^{2i\Delta\Omega_1}-1\big)\zeta(0)=-I,\qquad I\equiv\oint e^{2i\Omega_1}\Psi\,dt,
\label{eq:obstr}
\end{equation}
where $\Delta\Omega_1$ is the nodal advance per ZLK period. Off the commensurability
$\Delta\Omega_1=k\pi$ the solution $\zeta(0)=-I/(e^{2i\Delta\Omega_1}-1)$ is unique. On it,
a solution exists if and only if the obstruction integral $I$ vanishes.

\begin{figure*}[tb]
\centering
\includegraphics[width=\linewidth]{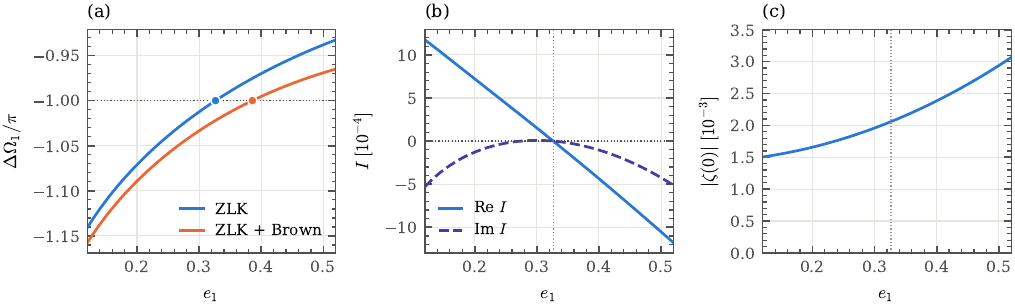}
\caption{(a) Nodal advance per ZLK period (the time in which $\omega_1$ advances by $\pi$)
for $j_z=0.4$, $\omega_1(0)=0.3$, $a_2/a_1=14$, $e_2=0.6$, $(m_1,m_2,m_3)=(1.5,0.5,7)$.
The commensurability $\Delta\Omega_1=-\pi$ lies at $e_1=0.3263$ for the ZLK flow and at
$0.3857$ with the Brown term added. (b) Real and imaginary parts of the obstruction
integral, Eq.~\eqref{eq:obstr}; both cross zero at the commensurability. (c)
$|\zeta(0)|=|I|/|e^{2i\Delta\Omega_1}-1|$, smooth through the commensurability.}
\label{fig:obstr}
\end{figure*}

The period must be determined with care. On a circulating ZLK orbit $e_1(\omega_1)$ is
symmetric about $\omega_1=\pi$, and $G_1$ returns to its initial value already at
$\omega_1=\pi-\omega_1(0)$, halfway through the period; the period is the time in which
$\omega_1$ advances by $\pi$. (For $j_z=0.4$, $e_1=0.2445$, $\omega_1(0)=0.3$ the
intermediate return occurs at $t=746$ and the period is $871$ in the units of
Fig.~\ref{fig:obstr}.) An integration over the intermediate arc gives a nonzero $I$ on
the would-be commensurability and would suggest, wrongly, that the harmonic is physical.

Over the full period we find, for $j_z=0.4$ and $\omega_1(0)=0.3$ in the test-particle
frame ($m_1,m_2,m_3=1.5,0.5,7$, $a_1=1$, $a_2=14$, $e_2=0.6$, $G=1$), the values listed in
Table~\ref{tab:obstr}.
$I$ crosses zero linearly at the commensurability [both real and imaginary parts,
Fig.~\ref{fig:obstr}(b)], and $\zeta(0)$ is continuous through it, varying smoothly from
$1.5\times10^{-3}$ to $3.1\times10^{-3}$ over $0.12\le e_1\le0.52$
[Fig.~\ref{fig:obstr}(c)]. The commensurabilities at $j_z=0.2$ ($e_1=0.365876$) and
$j_z=0.6$ ($e_1=0.253491$) give $|I|\sim10^{-14}$ as well. Eq.~\eqref{eq:gaugeeq}
is therefore solvable globally: \emph{the $2\omega_2$ harmonic is a gauge term}.

\begin{table}[tb]
\caption{Nodal advance per ZLK period, $\Delta\Omega_1$, and modulus of the obstruction integral $|I|$ over the full period, for $j_z=0.4$ and $\omega_1(0)=0.3$ in the test-particle frame ($m_1,m_2,m_3=1.5,0.5,7$, $a_1=1$, $a_2=14$, $e_2=0.6$, $G=1$). The column $e_1=0.326344$ is the commensurability $\Delta\Omega_1=-\pi$.}
\label{tab:obstr}
\centering\footnotesize\setlength{\tabcolsep}{2pt}
\begin{tabular}{@{}lccccc@{}}
\toprule
$e_1$ & 0.20 & 0.30 & 0.326344 & 0.35 & 0.40\\
\midrule
$\Delta\Omega_1/\pi$ & $-1.0710$ & $-1.0124$ & $-1.0000$ & $-0.9897$ & $-0.9700$\\
$|I|$ & $7.3{\times}10^{-4}$ & $1.5{\times}10^{-4}$ & $7{\times}10^{-15}$ & $1.4{\times}10^{-4}$ & $4.5{\times}10^{-4}$\\
\bottomrule
\end{tabular}
\end{table}

\subsection{The removing generator: fictitious time}
\label{sec:q2-tau}

The function $c$ can be written down explicitly. If the outer average is taken with a
fictitious time $\tau$ with $dt/d\tau$ a periodic function of $\M_2$ of unit mean
\cite{Tremaine2023}, the integrand of the first-order generator is unchanged, but the
integration constant is fixed by $\av{W}_\tau=0$ instead of $\av{W}_{\M_2}=0$:
\begin{gather}
W_\tau=W^*+\gamma_\tau,\qquad \gamma_\tau=-\av{W^*}_\tau,\nonumber\\
K^{(0,9/2)}_\tau=K^{(0,9/2)}+\{K_{\rm Q},\gamma_\tau\},
\label{eq:gamtau}
\end{gather}
by the gauge theorem, Eq.~\eqref{eq:gaugethm}. Let $b_k$ be the coefficients of $\sin kF$ in
$\bar Hw_2$. The even part of $\tilde G$ is $-\sum_k(b_k/k)\cos kF$, while the odd
functions ($F$, $\M_2$, $\sin kF$) average to zero with either measure, so that
\begin{equation}
\gamma_\tau=\frac1{n_2}\sum_k\frac{b_k}{k}\Big(\av{\cos kF}_\tau-\av{\cos kF}_{\M_2}\Big),
\label{eq:gamsum}
\end{equation}
with $\av{\cos kF}_{\M_2}=(1+k\ell_2)[-e_2/(1+\ell_2)]^k$. For the quadrupole
$b_1:b_2:b_3=e_2:2:e_2$ multiply one and the same angular function, and
\begin{equation}
\gamma_\tau=-\frac{3A_{\rm Q}}{4n_2}\,\Lambda_\tau(e_2)\,\mathcal S,
\label{eq:gamclosed}
\end{equation}
\begin{equation}
\begin{split}
\mathcal S={}&2\sin^2\iota\,(2+3e_1^2)\sin2\omega_2\\
&+5e_1^2\big[(1-\cos\iota)^2\sin(2\omega_1+2\omega_2)\\
&\qquad\ \ -(1+\cos\iota)^2\sin(2\omega_1-2\omega_2)\big],
\end{split}
\end{equation}
with $\Lambda_{\M}=0$ for the mean anomaly, and
\begin{equation}
\Lambda_F=\frac{e_2^2(1+2\ell_2)}{3(1+\ell_2)^2},\qquad
\Lambda_{F+e\sin F}=\Lambda_F+\frac{e_2^2}{2}
\label{eq:Lam}
\end{equation}
for the true anomaly and for $\tau=F+e_2\sin F$. The bracket $\{K_{\rm Q},\gamma_\tau\}$ has
the angular structure of $\mathcal Y_{92}$; its coefficient is
$-(1-\tfrac35/\Hf)$ times that of Eq.~\eqref{eq:K92closed} for $\tau=F$ and exactly $-1$
for $\tau=F+e_2\sin F$. In the latter gauge the $2\omega_2$ harmonic disappears and
$j_z$ is conserved, which is Tremaine's form $C_3=0$; for $\tau=F$ the ratio
$\tfrac35/\Hf$ is his $C_2/C_1$.

We verified Eq.~\eqref{eq:gamclosed} in two ways. The symbolic bracket
$\{K_{\rm Q},\gamma_\tau\}$ evaluated at a generic point agrees with the prediction of
Eq.~\eqref{eq:K92closed} to nine digits for both fictitious times. Independently, repeating
the numerical second-order calculation with $W^*\to W^*+\gamma_\tau$ multiplies all three
coefficients of the $2\omega_2$ harmonic by $0.651475$ $[=\tfrac35/\Hf(0.6)]$ for $\tau=F$
and by $0$ for $\tau=F+e_2\sin F$, leaving the Brown term unchanged to six digits. The
freedom found by Tremaine is therefore the gauge freedom of Eq.~\eqref{eq:gaugethm}, the same
mechanism that produces LR's Eq.~(4.5) from the inner 1PN precession
[Eq.~\eqref{eq:K32gauge}].

\subsection{Single-averaged test and the commensurability}
\label{sec:q2-sa}

\begin{figure*}[tb]
\centering
\includegraphics[width=\linewidth]{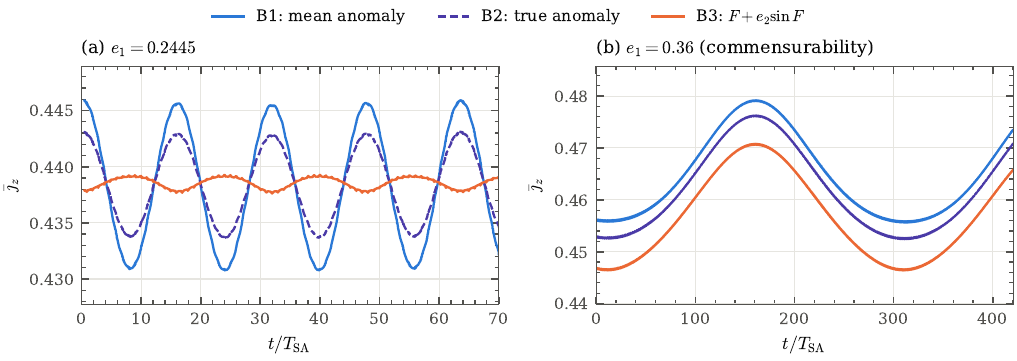}
\caption{Mean $j_z$ of one and the same single-averaged trajectory, mapped to mean elements
with the first-order map of three gauges, Eq.~\eqref{eq:jbar}: B1 (mean anomaly), B2 (true
anomaly), B3 ($F+e_2\sin F$); running mean over a ZLK period; $a_2/a_1=14$, $e_2=0.6$,
$j_z=0.4$, $\Omega_1(0)=0$. (a) $e_1=0.2445$, away from the commensurability: the
$2\omega_2$ oscillation of B1 and B2 disappears in B3. (b) $e_1=0.36$, on the
commensurability: the three gauges show the same slow variation, shifted by nearly
constant offsets.}
\label{fig:jbar}
\end{figure*}

The gauge statement can be tested dynamically. We integrate the single-averaged (SA)
quadrupole equations, in which $\bar H$ depends explicitly on the outer true anomaly
$F(t)$, with the explicit Runge--Kutta method DOP853 of Hairer, N{\o}rsett and Wanner
\cite{HairerNorsettWanner1993} (an eighth-order Dormand--Prince scheme with embedded
error estimators of orders five and three and dense output), as implemented in
SciPy \cite{SciPy2020} (relative tolerance $10^{-10}$, maximum step $P_{\rm out}/16$), and map the
osculating state to mean elements with the first-order map of each gauge,
\begin{equation}
\bar\jmath_z^{(\tau)}=\frac1{L_1}\Big[H_1-\frac{\partial}{\partial\omega_2}
\big(W^*+\gamma_\tau\big)\Big],
\label{eq:jbar}
\end{equation}
for $\tau=\M_2$ (B1), $F$ (B2) and $F+e_2\sin F$ (B3). The double-averaged dynamics
predicts an oscillating $\bar\jmath_z$ in B1 and B2 and a constant one in B3.
Figure~\ref{fig:jbar}(a) shows the slow part (running mean over a ZLK period) for
$e_1=0.2445$, $j_z=0.4$: its peak-to-peak variation is $1.51\times10^{-2}$ in B1,
$0.94\times10^{-2}$ in B2 and $0.17\times10^{-2}$ in B3, against the double-averaged
predictions $1.49\times10^{-2}$, $0.98\times10^{-2}$ and $0$. The residual in B3 is of
order $\varepsilon_{21}^2$, beyond the first-order map. The same trajectory thus
appears to change $j_z$ or to conserve it, depending on the definition of the mean elements.

In the B1 elements the slow variation has the period $\pi T_{\rm SA}/|\delta|$, with
$\delta=\Delta\Omega_1+\pi$ the detuning from the commensurability measured on the SA mean
orbit: this is the rotation of the phase of the gauge oscillation
$c=\mathrm{Re}[\zeta e^{2i\omega_2}]$ from one ZLK period to the next. For $e_1=0.2445$
the measured $\Delta\Omega_1/\pi=-1.0628$ predicts $15.94\,T_{\rm SA}$, and the SA
integration gives $15.86\,T_{\rm SA}$.

On the commensurability the picture changes [Fig.~\ref{fig:jbar}(b), $e_1=0.36$ in
osculating elements, whose SA mean orbit $(\bar e_1,\bar\jmath_z)=(0.3128,0.4527)$ satisfies
$\Delta\Omega_1/\pi=-1.0028$]. All three gauges now show the same slow variation, with a
peak-to-peak amplitude of $2.4\times10^{-2}$ and a period longer than $300\,T_{\rm SA}$,
while their differences are nearly constant (B1$-$B3 varies by $2.4\times10^{-3}$ about a
mean of $0.008$). The differences are constant because on the commensurability the phase
of $e^{2i\Omega_1}\zeta$ does not rotate from one period to the next; the common variation
is independent of the gauge and hence dynamical. It cannot be driven by the
quadrupole-squared harmonic, whose obstruction integral vanishes there, and must come from
the next order, $\mathcal{O}(\varepsilon_{21}^2)$ relative to the ZLK term (the cube of the
quadrupole), whose $2\omega_2$ part need not be a gauge term. Such a second-order
resonance would have an amplitude $\propto\varepsilon_{21}$ and a period
$\propto T/\varepsilon_{21}$; we leave its analysis to future work.

\subsection{The inclination chain rule at quadrupole-squared order}
\label{sec:q2-iota}

Paper~I found that LR's implementation breaks the chain rule for
$\iota=\iota_1+\iota_2$, so that the re-substitution term in the $\iota$ channel,
$\Delta_\alpha=\av{(\partial Q_\alpha/\partial\iota)W_\iota}_t$, is lost ($Q_\alpha$ is the
quadrupole rate of element $\alpha$ as a function of the outer phase and $W_\iota$ the
quadrupole periodic solution for $\iota$), and that this term vanishes for the
1PN\,$\times$\,quadrupole cross terms but not at quadrupole-squared order. We have evaluated
it in the test-particle limit (the regime of LR) and compared it with the correct
second-order field $X_{K_2}$ of $K_2=\frac12\av{\{H-\av H,W_H\}}_t$, whose $\omega_2$ average
agrees with the Brown Hamiltonian to five digits. Averaged over $\omega_2$, the ratio
$\Delta_\alpha/(X_{K_2})_\alpha$ is given in Table~\ref{tab:iota}.
The lost terms are of the same magnitude as the Brown rates. Moreover, whereas the
$\omega_2$-averaged Brown field conserves $j_z$, the lost terms do not; for $a_2/a_1=14$,
$\varepsilon_{21}\simeq0.08$ they would produce a spurious drift
$|\Delta j_z|\sim10^{-3}$--$10^{-2}$ per ZLK period. An extension of the multiple-scale
implementation to quadrupole-squared order would therefore have to restore the $\iota$
channel.

\begin{table}[tb]
\caption{Ratio of the terms lost in the $\iota$ substitution, $\Delta_\alpha$, to the correct second-order field $(X_{K_2})_\alpha$, averaged over $\omega_2$, in the test-particle limit for four sample states $(e_1,\iota,\omega_1,e_2)$.}
\label{tab:iota}
\centering\small
\begin{tabular}{@{}lcccc@{}}
\toprule
$(e_1,\iota,\omega_1,e_2)$ & $\dot e_1$ & $\dot\iota$ & $\dot\omega_1$ & $\dot\Omega_1$\\
\midrule
$(0.3,0.9,0.3,0.6)$ & $-0.42$ & $0.38$ & $0.013$ & $0.40$\\
$(0.6,1.2,1.1,0.6)$ & $-0.80$ & $1.74$ & $0.41$ & $0.42$\\
$(0.2,0.6,2.0,0.3)$ & $-0.37$ & $0.08$ & $0.067$ & $0.72$\\
$(0.5,1.4,0.7,0.5)$ & $-0.61$ & $11.1$ & $0.17$ & $0.44$\\
\bottomrule
\end{tabular}
\end{table}

\section{Orbital evolution: comparison of the formulations}
\label{sec:orbit}

We now compare the secular evolution given by the three sets of cross terms for the triples
that LR used to illustrate them (their Figs.~3--8), with the masses
$(m_1,m_2,m_3)=(30,20,2\times10^7)\,M_\odot$ and $e_2=0.8$. We use the secular code of
Paper~I (orbit-averaged equations for the angular-momentum and eccentricity vectors of both
orbits, integrated with the embedded Runge--Kutta Prince--Dormand (8,9) stepper
\cite{PrinceDormand1981} of the GNU Scientific Library \cite{GSL2009} with tolerances
$10^{-12}$), in which the canonical
terms of this paper were added as a Hamiltonian $K^{(1,3)}+K^{(1,4)}_{\rm quad}$ acting on the
inner orbit through the Milankovitch equations \cite{Milankovitch1939,AllanWard1963,TremaineToumaNamouni2009},
Eq.~\eqref{eq:milank}; the gradients of $K$ with respect to the inner vectors are taken by
central differences.
The implementation reproduces $\dot G_1=-\partial K/\partial\omega_1$ of the symbolic result to
$10^{-3}$ (the difference reflects the values of the physical constants). The four models are
listed in Table~\ref{tab:models}. All include the Newtonian quadrupole (and, where stated, the
octupole) and the two-body 1PN precessions of both orbits.

\begin{table}[tb]
\caption{Models compared in Sec.~\ref{sec:orbit}.}
\label{tab:models}
\centering\small
\begin{tabular}{@{}>{\raggedright\arraybackslash}p{0.2\columnwidth}>{\raggedright\arraybackslash}p{0.74\columnwidth}@{}}
\toprule
Model & Cross terms\\
\midrule
no 3BpN & none\\
LR & LR's Eqs.~(4.4), (4.5), (4.6) and the other $\delta\varepsilon^4$ terms\\
Paper I & as LR without Eq.~(4.5), with KST's magnetic-quadrupole term\\
this work & de Sitter [Eq.~\eqref{eq:KdS}], electric and next-to-leading order
1PN\,$\times$\,quadrupole [Eqs.~\eqref{eq:K13}, \eqref{eq:K14}], magnetic quadrupole
[Eq.~\eqref{eq:K72}]; no $\delta\varepsilon^{3/2}$ term\\
\bottomrule
\end{tabular}
\end{table}

\subsection{Librating and circulating systems}

\begin{figure*}[tb]
\centering
\includegraphics[width=\linewidth]{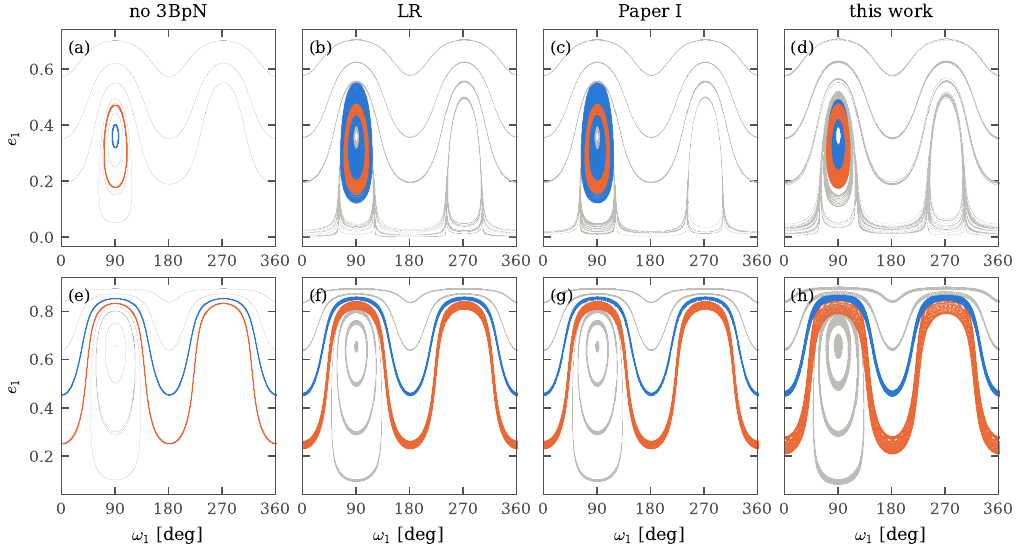}
\caption{Trajectories in the $(\omega_1,e_1)$ plane over 2000~yr for LR's librating system
[top, LR Fig.~3: $a_1=0.10$~AU, $a_2=209.84$~AU, $l_z=-0.6593$] and circulating system
[bottom, LR Fig.~5: $a_1=0.94$~AU, $a_2=191.86$~AU, $l_z=0.4449$], for the four models of
Table~\ref{tab:models}. The initial $e_1$ varies at fixed $l_z=\ell_1\cos\iota_1$; blue: initial
$e_1=0.40$ (top) and $0.85$ (bottom); orange: $0.47$ and $0.83$, close to the separatrix.}
\label{fig:maps}
\end{figure*}

\begin{figure}[tb]
\centering
\includegraphics[width=\columnwidth]{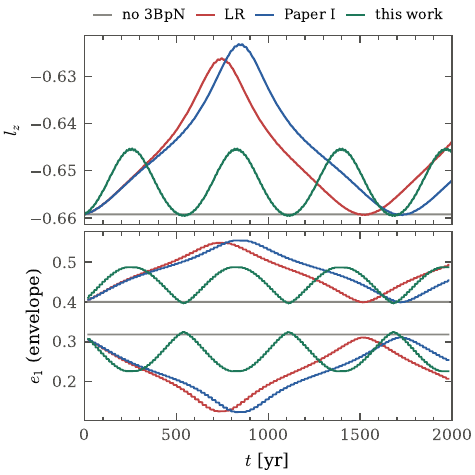}
\caption{Time evolution of $l_z$ (top) and of the envelope of $e_1$ (running maximum and
minimum over 40~yr, bottom) for the blue librating trajectory of Fig.~\ref{fig:maps}(a)--(d).}
\label{fig:evol}
\end{figure}

Figure~\ref{fig:maps} repeats Fig.~3 of Paper~I with the canonical model added. Without cross
terms $l_z$ is conserved and the trajectories are closed. With the cross terms $l_z$ is
modulated and the trajectories thicken. The LR and Paper~I models differ only through Eq.~(4.5),
which changes the period of the modulation but not its envelope (Paper~I). The
canonical model differs from both. For the librating trajectory with initial $e_1=0.40$, $e_1$
ranges over $[0.125,0.548]$ (LR), $[0.123,0.555]$ (Paper~I) and $[0.226,0.487]$ (this work), and
the maximum of $l_z$ is $-0.6262$, $-0.6232$ and $-0.6453$; the period of the $l_z$ modulation is
shorter in the canonical model, whose first maximum of $l_z$ occurs at 258~yr instead of 745~yr
(LR) and 838~yr (Paper~I) (Fig.~\ref{fig:evol}). For the circulating trajectories the canonical
model gives the wider band: for initial $e_1=0.85$ the range of $l_z$ is $[0.4185,0.4530]$
against $[0.4330,0.4487]$ in LR, and for $e_1=0.83$ it is $[0.4057,0.4884]$ against
$[0.4331,0.4685]$.

To identify the responsible terms we switched the canonical terms on one at a time, keeping the
de Sitter term. With the electric term alone, or with only the ZLK-like part $c_ZZ$ of the
next-to-leading order term, the trajectories are indistinguishable from those without cross terms
[for the librating trajectory $e_1\in[0.318,0.400]$ and $[0.320,0.400]$, respectively, against
$[0.318,0.400]$ without cross terms, and $l_z$ constant to $10^{-5}$]: these parts only renormalize the ZLK precession rate. The modulation is produced
entirely by the $\omega_2$-dependent part $c_YY$ of the next-to-leading order term; alone it gives
$e_1\in[0.194,0.511]$ and $\max l_z=-0.6386$, and the ZLK-like part reduces this to the full
result. The difference between the LR (or Paper~I) model and ours is therefore the difference
between LR's $\omega_2$-dependent coefficient in Eq.~(4.6) and our $c_Y$, discussed in
Sec.~\ref{sec:pnq-n}: ours contains the de Sitter\,$\times$\,ZLK cross term, and LR's is half of
a gauge-shifted outer-1PN\,$\times$\,ZLK term. The de Sitter term alone does not modulate $l_z$,
because it depends on $\iota$ but not on $\omega_2$.

\subsection{The accelerated merger of LR Fig.~8}

\begin{figure*}[tb]
\centering
\includegraphics[width=0.86\linewidth]{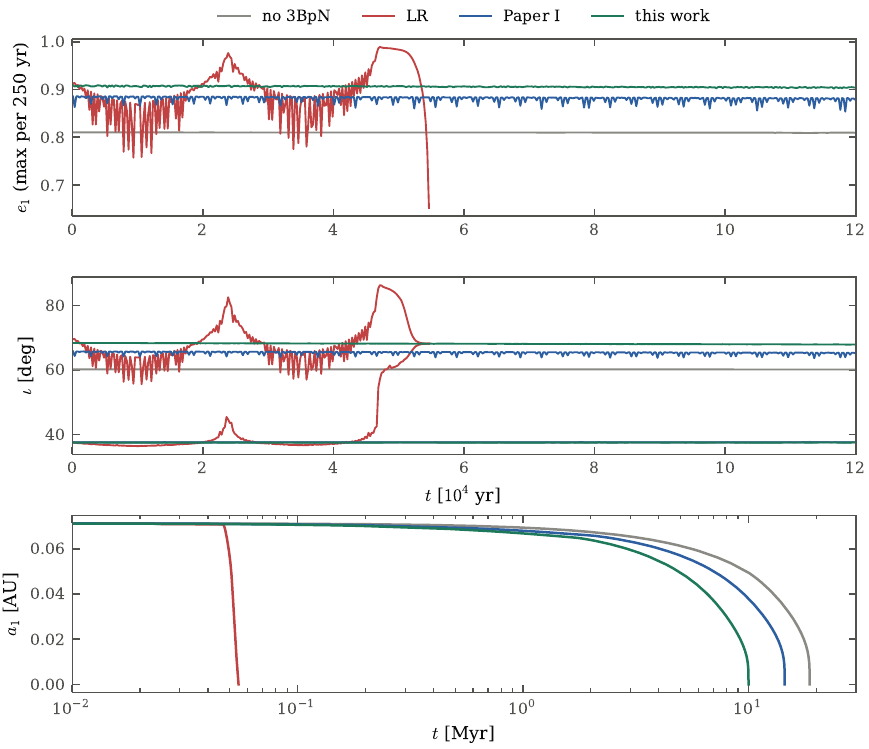}
\caption{Evolution with octupole terms and gravitational-wave emission for LR's Fig.~8
system in orientation A of Paper~I ($a_1=0.0713$~AU, $e_1=0.808$, $a_2=144.694$~AU,
$\iota=37.935^\circ$). Top two panels: range of $e_1$ and $\iota$ in bins of 250~yr over the
first $1.2\times10^5$~yr; bottom: $a_1$ until merger.}
\label{fig:merger}
\end{figure*}

LR's Fig.~8 shows a triple in which the cross terms, together with the octupole, raise
$e_{1,\max}$ from $0.81$ to $0.99$ and shorten the merger time by a factor of about 36.
Paper~I found that this acceleration is produced by Eq.~(4.5). Figure~\ref{fig:merger} adds the
canonical model for the orientation of Paper~I (orientation A). The merger times are
$1.87\times10^7$~yr without cross terms, $5.5\times10^4$~yr with LR's terms,
$1.45\times10^7$~yr in the model of Paper~I and $1.00\times10^7$~yr in the canonical model, with
$e_{1,\max}=0.810$, $0.986$, $0.886$ and $0.909$. The canonical cross terms therefore raise the
maximum eccentricity and shorten the merger time relative to Paper~I, but do not produce the
large transitions of $e_1$ and $\iota$ ($\iota>80^\circ$) that precede the rapid merger in the LR
model. Table~\ref{tab:robust} shows the same pattern for the three orientations of Paper~I.

\begin{table}[tb]
\caption{Maximum eccentricity over $3\times10^5$~yr (octupole included, no gravitational-wave
emission) for the three orientations A, B, C of Paper~I at the parameters of
Fig.~\ref{fig:merger}.}
\label{tab:robust}
\centering\small
\begin{tabular}{@{}lccc@{}}
\toprule
Model & A & B & C\\
\midrule
no 3BpN & 0.8104 & 0.8165 & 0.8130\\
LR & 0.9869 & 0.9786 & 0.9782\\
Paper I & 0.8862 & 0.8195 & 0.8587\\
this work & 0.9092 & 0.8245 & 0.8839\\
\quad electric term only & 0.8126 & & \\
\quad $c_ZZ$ only & 0.8124 & & \\
\quad $c_YY$ only & 0.9067 & & \\
\bottomrule
\end{tabular}
\end{table}

As for the systems of Fig.~\ref{fig:maps}, switching the canonical terms on separately (last three
rows) shows that the increase of $e_{1,\max}$ comes from the $\omega_2$-dependent part $c_YY$ of
the next-to-leading order term; the electric term and the ZLK-like part change $e_{1,\max}$ by only
$2\times10^{-3}$.

\subsection{A sample of orientations}

\begin{figure*}[tb]
\centering
\includegraphics[width=\linewidth]{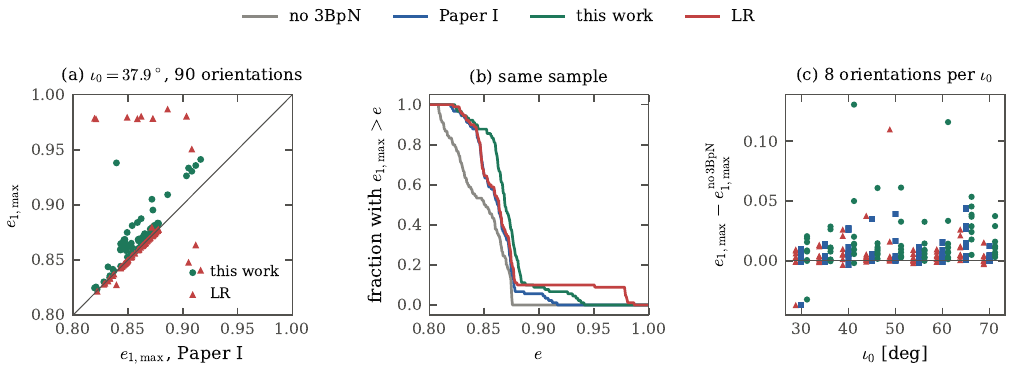}
\caption{Maximum eccentricity over $3\times10^5$~yr at the parameters of Fig.~\ref{fig:merger}
(octupole included, no gravitational-wave emission). (a) The canonical model (circles) and LR
(triangles) against the model of Paper~I, for 90 random orientations at $\iota_0=37.9^\circ$.
(b) Cumulative distributions of the same sample. (c) Increase of $e_{1,\max}$ over the model
without cross terms for 8 orientations at each $\iota_0$.}
\label{fig:pop}
\end{figure*}

We repeated the orientation sample of Paper~I with the canonical model (Fig.~\ref{fig:pop}).
Among 90 orientations at $\iota_0=37.9^\circ$, $e_{1,\max}>0.95$ is reached by none of the
orientations without cross terms, by 9 in the LR model, and by none in the Paper~I and canonical
models; $e_{1,\max}>0.9$ is reached by 0, 9, 5 and 8, and the largest values are $0.875$,
$0.987$, $0.916$ and $0.941$, respectively. The canonical model raises $e_{1,\max}$ above the value
without cross terms for 94\% of the orientations; the median increase is $0.011$, against
$0.004$ for Paper~I. The effect is systematic in inclination [Fig.~\ref{fig:pop}(c)]: averaged over
the eight orientations at each $\iota_0$ between $30^\circ$ and $70^\circ$, the increase in the
canonical model is larger than that of Paper~I at every $\iota_0$, by factors of 1.7--5.2 for
$\iota_0\ge35^\circ$ (for example $0.030$ against $0.010$ at $\iota_0=40^\circ$), and at $\iota_0=60^\circ$ two orientations exceed $0.95$ in the canonical model
and none in the others. Table~\ref{tab:robust} lists the three orientations of Paper~I.

\subsection{Summary of the comparison}

In the triples of LR the relativistic modulation of $l_z$ and the enhancement of the maximum
eccentricity with the octupole are controlled by a single term, the $\omega_2$-dependent part of the
next-to-leading order 1PN\,$\times$\,quadrupole Hamiltonian; the electric and magnetic
1PN\,$\times$\,quadrupole terms and the ZLK-like part of the next-to-leading order term are negligible
there. LR's Eq.~(4.6) contains a term of this structure with a different coefficient (it lacks the
de Sitter\,$\times$\,ZLK contribution), so that the envelopes and periods of the modulations differ
between LR and the canonical theory. With the canonical coefficient the modulations are faster and,
for librating systems, of smaller amplitude, while the increase of $e_{1,\max}$ with the octupole is
two to three times as large as in the model of Paper~I; the accelerated mergers of LR, which require
Eq.~(4.5), remain absent. Whether the canonical enhancement changes the population statistics of LR
is the subject of Paper~III.

\section{Discussion}
\label{sec:disc}

\subsection{Relation to previous work}

Table~\ref{tab:prev} collects the comparison order by order. At the two orders where
Paper~I found the formulations in disagreement, the canonical calculation agrees with KST:
the order $\delta\varepsilon^{3/2}$ is empty, and the magnetic-quadrupole term is $\Kmq$ of
Eq.~\eqref{eq:Kmq}. At the orders on which the formulations had not been compared directly,
the electric 1PN\,$\times$\,quadrupole term agrees exactly with KST; the next-to-leading order term,
which KST do not give (it is beyond their order), differs from LR's Eq.~(4.6) both in the
$\omega_2$-dependent part (by a factor that depends on $e_2$) and in the
$\omega_2$-independent part (by sign); the latter can be changed by a redefinition of the
elements, the former is suggestive of a measure mismatch of the kind identified in Paper~I.
LWYL use the same grading and report that the direct term at order $\delta\varepsilon^{1/2}$
vanishes, that the order $\delta\varepsilon$ only shifts the periods, that the order
$\delta\varepsilon^2$ interaction term is removed by their canonical transformation, and that
the cross terms at orders $\delta\varepsilon^{3/2}$ and $\delta\varepsilon^{5/2}$ average to zero;
all of this agrees with Sec.~\ref{sec:mono}. A term-by-term comparison of their
1PN\,$\times$\,quadrupole expressions with ours requires their intermediate results
(their generator of the second transformation has a nonzero mean, i.e.\ $\gamma\neq0$ in
our language), and we have not attempted it.

\begin{table*}[tb]
\caption{Comparison with previous work, order by order.}
\label{tab:prev}
\centering\small
\begin{tabular}{@{}p{0.15\textwidth}p{0.06\textwidth}p{0.15\textwidth}p{0.14\textwidth}p{0.12\textwidth}p{0.28\textwidth}@{}}
\toprule
Term & Order & LR & KST & LWYL & This work\\
\midrule
Monopole, odd & $(1,\tfrac12)$ & -- & -- & direct term $=0$ & $K=0$, Eq.~\eqref{eq:K12}\\
Binding energy & $(1,1)$ & -- & -- & periods only & mass renormalization, Eq.~\eqref{eq:renorm}\\
Libration & $(1,\tfrac32)$ & Eq.~(4.5) & absent & averages to zero & $H\equiv0$; gauge term in the $F$ measure\\
Outer 1PN, dipole & $(1,2)$ & -- & -- & removed & two-body 1PN; dipole averages to zero\\
de Sitter & $(1,\tfrac52)$ & Eq.~(4.4) & agree & agree & agree, Eq.~\eqref{eq:KdS}\\
Electric 1PN$\times$Q & $(1,3)$ & (A2)--(A6) & Eq.~(35), 1st term & -- & $=$ KST exactly, Eq.~\eqref{eq:K13}\\
Magnetic 1PN$\times$Q & $(1,\tfrac72)$ & (A7)--(A11), disagree & Eq.~(35), 2nd term & -- & $=$ KST exactly, Eq.~\eqref{eq:K72}\\
Next-to-leading order 1PN$\times$Q & $(1,4)$ & Eq.~(4.6) etc., differ & -- & -- & Eqs.~\eqref{eq:K14}--\eqref{eq:cY}\\
Quadrupole-squared & $(0,\tfrac92)$ & -- & -- & -- & $=$ Will's secular equations; $2\omega_2$ harmonic is gauge\\
\bottomrule
\end{tabular}
\end{table*}

For the quadrupole-squared term the situation is now clear. Our Hamiltonian in the time gauge
generates the secular equations of Will \cite{Will2021} and Conway and Will
\cite{ConwayWill2024}; the Brown Hamiltonian of Refs.~\cite{Tremaine2023,LeiGrishin2025a} is
the same dynamics in the gauge $\tau=F+e_2\sin F$; and the correction potential of Luo, Katz,
and Dong \cite{LuoKatzDong2016}, which depends on the direction of the outer eccentricity
vector, belongs to the same family (Will's Appendix~C relates the two by an algebraic change of
elements). The obstruction integral shows that these descriptions are equivalent everywhere,
including the commensurability of the ZLK period with the nodal precession, and the generator
$\gamma_\tau$ makes the equivalence explicit. The dynamics near that commensurability is not
empty, however: the single-averaged integration reveals a gauge-independent slow variation of
$j_z$ that must be attributed to the next order.

\subsection{Gauge in the three formulations}

The word ``gauge'' has entered this series in three roles, which it may be useful to
separate. (i) The EIH and ADM formulations share the same 1PN Hamiltonian; the
difference between osculating and contact elements is a near-identity map that includes the
tertiary-induced rescaling $3Gm_3/(c^2R)$ of the velocity and changes secular rates first at
order $\delta\varepsilon^4$ in the part induced by the tertiary. (ii) The normalization of the
periodic part of the generator (the averaging measure) changes the second-order secular
Hamiltonian by $\{K_1,\gamma\}$; LR's Eq.~(4.5) and the $2\omega_2$ harmonic of the
quadrupole-squared term are examples. (iii) Physical conclusions --- the maximum eccentricity,
the merger time, whether a trajectory librates --- are independent of (i) and (ii), and
comparisons of secular rates between formulations are meaningful only after the elements have
been brought to a common definition, or through gauge-invariant properties such as the
existence of a secular Hamiltonian (Paper~I) or the vanishing of an obstruction integral
(Sec.~\ref{sec:q2-gauge}).

\subsection{Limitations and outlook}

The canonical terms were implemented in the secular code as a Hamiltonian acting on the inner
orbit only, which is adequate for $m\ll m_3$; for comparable masses the back-reaction on the
outer orbit follows from the same Hamiltonian and should be included. The octupole-type part of
the next-to-leading order 1PN\,$\times$\,quadrupole term is not yet implemented; it is suppressed by
$e_2|m_1-m_2|/m$ relative to the quadrupole-type part but is not negligible for unequal masses.
The next Newtonian order, the cube of the quadrupole and the square of the octupole, is needed to
describe the slow $j_z$ variation found at the commensurability. Finally, the population study
of LR, in which the cross terms raise the maximum eccentricity of a large fraction of
triples, will be recomputed with the canonical secular Hamiltonian in Paper~III.

\section{Summary}
\label{sec:sum}

We have derived the secular Hamiltonian of hierarchical triples to 1PN\,$\times$\,quadrupole
and quadrupole-squared order from the ADM Hamiltonian by a two-stage Lie--Hori--Deprit
transformation in canonical variables, enumerating all contributions. The main results are
the following.
\begin{enumerate}[label=(\arabic*),leftmargin=16pt]
\item The EIH and ADM formulations share the same 1PN Hamiltonian; they differ in the
      definition of the elements. The osculating--contact map contains a tertiary-induced
      term of relative magnitude $3Gm_3/(c^2R)$, Eq.~\eqref{eq:oscon-m3}.
\item The 1PN monopole interaction renormalizes the mass of the inner binary by its binding
      energy, equally in the inertial and gravitational mass; in PPN form the gravitational
      mass acquires the Nordtvedt term $\eta_N\Omega_b/c^2$. The order $\delta\varepsilon^{3/2}$
      is empty.
\item The electric and magnetic 1PN\,$\times$\,quadrupole cross terms agree exactly with KST;
      the magnetic one is shared between a direct term and a second-stage cross term in the
      ratio $(4m+3m_3):m_3$.
\item The next-to-leading order 1PN\,$\times$\,quadrupole term, Eqs.~\eqref{eq:K14}--\eqref{eq:cY},
      depends on $\omega_2$ through the outer eccentricity; it contains a
      de Sitter\,$\times$\,ZLK cross term absent from LR.
\item The quadrupole-squared Hamiltonian, Eq.~\eqref{eq:K92closed}, generates the secular
      equations of Will and of Conway and Will; its $2\omega_2$ harmonic is a gauge term, removed
      by the generator $\gamma_\tau$ of Eq.~\eqref{eq:gamclosed}, which reproduces Tremaine's
      family of Brown Hamiltonians. At the commensurability of the ZLK period with the nodal
      precession a gauge-independent slow variation of $j_z$ remains, of higher order.
\item In the secular evolution of LR's triples the modulation of $l_z$ and the enhancement of
      $e_{1,\max}$ with the octupole are controlled by the $\omega_2$-dependent part of the
      next-to-leading order 1PN\,$\times$\,quadrupole term. With the canonical coefficient the
      modulations are faster and the enhancement is two to three times that of Paper~I
      (merger time $1.0\times10^7$~yr instead of $1.4\times10^7$~yr for LR's Fig.~8 system), while
      the accelerated mergers produced by LR's Eq.~(4.5) remain absent (Sec.~\ref{sec:orbit}).
\end{enumerate}

\begin{acknowledgments}
We are grateful to Halston Lim and Carl L. Rodriguez for kindly providing their Mathematica
notebooks and C++ code \cite{LimRodriguezPC}. This work was supported by the Research Grant of
the College of Engineering, Nihon University. Symbolic computations were carried out with
Wolfram Mathematica and SymPy, and numerical calculations with NumPy, SciPy and a modified
version of the C++ code of Lim and Rodriguez.
\end{acknowledgments}

\section*{Data availability}
The data that support the findings of this article are openly available~\cite{ArakidaCode2026,ArakidaKozaiFork}.
The Mathematica, Python and C++ code used for the symbolic derivations, the numerical checks and
the orbit integrations, with the expected outputs, is archived in Ref.~\cite{ArakidaCode2026}; the
modified C++ code, with its changes with respect to the code of Lim and Rodriguez recorded commit by
commit, is also available as a fork of the original repository~\cite{ArakidaKozaiFork}.

\appendix
\section{von Zeipel and Lie--Hori--Deprit transformations}
\label{app:vz}

LWYL use a von Zeipel transformation, whereas we use the Lie--Hori--Deprit method. We
show that the two give the same secular Hamiltonian when the same normalization is
imposed. Let $q=(\M,g)$ and $p=(L,G)$ collect a fast angle and slow angles with their
actions, $H=H_0(L)+\epsilon H_1+\epsilon^2H_2+\cdots$, $n=\partial H_0/\partial L$,
$\Delta=H_1-\av{H_1}$ and $\mathcal A$ the mean-free antiderivative in $\M$.

\paragraph*{von Zeipel.}
A generating function of mixed variables,
$S(q,P)=q\cdot P+\epsilon S_1(q,P)+\epsilon^2S_2+\cdots$, defines the map implicitly by
$p=\partial_qS$, $Q=\partial_PS$ \cite{Brouwer1959}. Expanding
$H(q,\partial_qS)=K(\partial_PS,P)$ order by order gives
$n\,\partial_{\M}S_1=K_1-H_1$, hence $K_1=\av{H_1}$ and
$S_1=-\mathcal A[\Delta]/n+\sigma_1(g,P)$, and
\begin{equation}
K_2^{\rm vZ}=\Big\langle H_2+\partial_PH_1\cdot\partial_qS_1+\tfrac12\partial_Ln\,(\partial_{\M}S_1)^2
-\partial_gK_1\,\partial_GS_1\Big\rangle .
\label{eq:vz2}
\end{equation}

\paragraph*{Lie--Hori--Deprit.}
The generator $W$ is a function on phase space and the map
$x^{\rm old}=\exp(\mathcal L_W)x^{\rm new}$ is explicit \cite{Hori1966}. Deprit's form
\cite{Deprit1969} defines the map as a flow in $\epsilon$ and yields the triangular
recursion $H^{(i)}_j=H^{(i-1)}_{j+1}+\sum_{k=0}^{j}\binom{j}{k}\{H^{(i-1)}_{j-k},W_{k+1}\}$,
$K_n=H^{(n)}_0$; the two forms differ in the normalization of $W_2$ but give the same $K$
\cite{CampbellJefferys1970,Kamel1969,Mersman1970}. The second-order result is
Eq.~\eqref{eq:K2avg}.

\paragraph*{Correspondence.}
Expanding both maps to second order gives
\begin{equation}
S_1=-W_1,\qquad S_2=-W_2+\tfrac12\,\partial_qW_1\cdot\partial_pW_1,
\label{eq:SW}
\end{equation}
which we verified symbolically with polynomial test generators. With
$\sigma_1=\gamma_1=0$ the second-order Hamiltonians agree. In the fast channel the von
Zeipel terms give $-\av{\Delta\,\partial_L\Delta}/n+\partial_Ln\av{\Delta^2}/(2n^2)$, and
the Poisson-bracket term $\tfrac12\av{\partial_{\M}\Delta\,\partial_LW_1-\partial_L\Delta\,\partial_{\M}W_1}$
gives the same after an integration by parts in $\M$. In the slow channel both give
$-\av{\partial_G\Delta\,\partial_gW_1}$, using $\Delta=n\,\partial_{\M}W_1$ and the independence
of $n$ from $G$. Hence $K_2^{\rm vZ}=K_2$ for any $n(L)$ and any dependence of $H_1$ on
$(\M,g)$; we also checked the identity symbolically for
$H_0=-1/(2L^2)$ and $H_1=L^2G\cos\M+(G^2/L)\cos(\M+2g)+LG^3\cos g+(L^3/G)\cos(2\M-g)$. The
same integration by parts, applied to two perturbations $a$ and $b$, proves the symmetry of
the bilinear form used in Eq.~\eqref{eq:cross}.

\paragraph*{Gauge.}
In both methods the first-order generator is defined only up to a function of the slow
variables ($\sigma_1=-\gamma_1$), and in both $K_2\to K_2+\{K_1,\gamma_1\}$. The time-mean-free
choice is Brouwer's; any other choice produces the gauge terms discussed in
Secs.~\ref{sec:lie-gauge} and \ref{sec:q2-tau}.

\paragraph*{Practical differences.}
The Lie method is explicit in both directions, applies to any function by
$F\mapsto\exp(\mathcal L_W)F$, uses only Poisson brackets and hence works equally with
Delaunay, nonsingular or vector (Milankovitch) elements \cite{Milankovitch1939,AllanWard1963,RosengrenScheeres2014}, and extends mechanically to higher order through
the triangular recursion. For the present problem it has three specific advantages: the two
stages and the several channels (inner, outer, slow, fast) are handled by single brackets,
which makes the transfer rules of Table~\ref{tab:transfer} immediate; the node elimination
$\partial_{\Omega_1}=-\partial_{\omega_2}$ and the abstract orientation algebra act directly on
$W$; and the map between contact and mean elements is explicit, as used in
Eq.~\eqref{eq:jbar}. The multiple-scale method \cite{Nayfeh1973,KevorkianCole1996} gives the
same first-order periodic solution, $\{X,W_1\}$, and the same second-order secular equations
provided the periodic solutions are time-mean-free and all channels of the chain rule are
retained \cite{SandersVerhulstMurdock2007}; the two problems identified in Paper~I are
violations of these two conditions.

\section{Kepler averages}
\label{app:kepler}

With $\av{\cdot}=(2\pi)^{-1}\oint(\cdot)\,d\M$ over a Kepler orbit of semimajor axis $a$,
eccentricity $e$, $\ell=\sqrt{1-e^2}$ and total mass $\mathfrak m$
($\hat{\bm\alpha}$ is the pericenter direction),
\begin{align}
&\av{r^{-1}}=\frac1a,\quad \av{r^{-2}}=\frac{1}{a^2\ell},\quad
\av{r^{-3}}=\frac{1}{a^3\ell^3},\nonumber\\
&\av{\bm r}=-\frac32ae\,\hat{\bm\alpha},\nonumber\\
&\av{v^2}=\frac{G\mathfrak m}{a},\quad
\av{v^4}=\Big(\frac{G\mathfrak m}{a}\Big)^2\Big(\frac4\ell-3\Big),\nonumber\\
&\av{\frac{v^2}{r}}=\frac{G\mathfrak m}{a^2}\Big(\frac2\ell-1\Big),\quad
\av{\frac{\dot r^2}{r}}=\frac{G\mathfrak m}{a^2}\Big(\frac1\ell-1\Big),\nonumber\\
&\av{v^iv^j}=G\mathfrak m\av{\frac{n^in^j}{r}}\quad\text{(tensor virial)},
\end{align}
and, for the true anomaly $f$,
\begin{equation}
\av{\cos kf}=(1+k\ell)\Big(-\frac{e}{1+\ell}\Big)^k,\qquad \av{\sin kf}=0 .
\end{equation}
For the outer orbit,
$\av{R^{-3}}=a_2^{-3}\ell_2^{-3}$ and
$\av{R^iR^j/R^5}=(\delta^{ij}-\hat z_2^i\hat z_2^j)/(2a_2^3\ell_2^3)$. The measures are
$d\M=(1-e\cos E)\,dE=\ell^3(1+e\cos f)^{-2}df$. All averages were checked numerically
to $10^{-15}$.

\section{Symbolic and numerical pipelines}
\label{app:pipe}

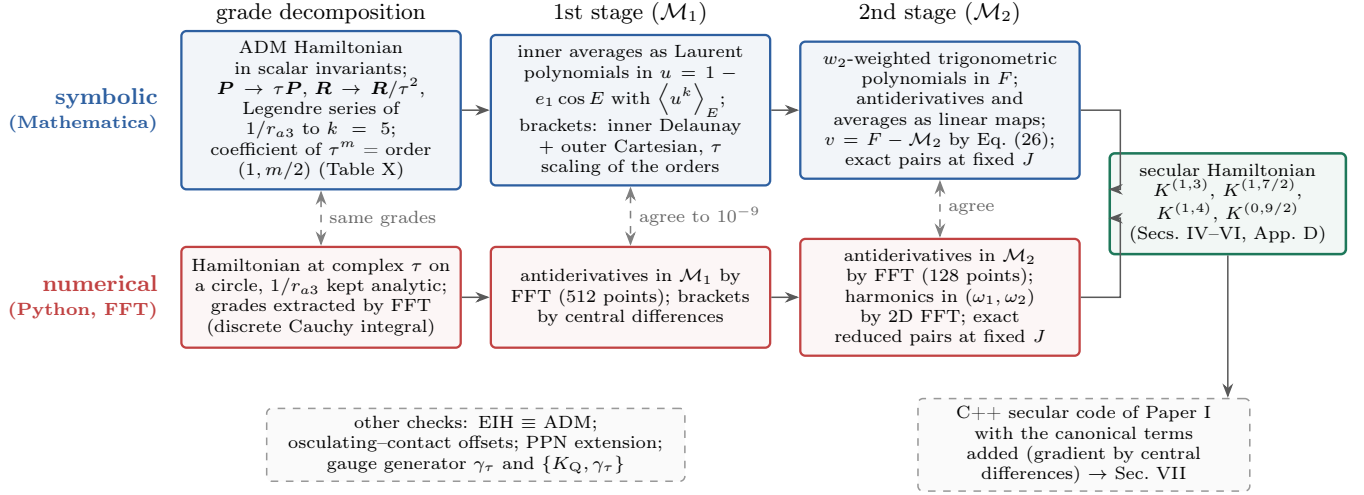
\begin{figure*}[tb]
\centering
\resizebox{\linewidth}{!}{\begin{tikzpicture}[>=Stealth,line width=0.6pt,font=\small,x=1cm,y=1cm,
 bx/.style={draw=black!55,rounded corners=2pt,align=center,execute at begin node=\nohyII,
       inner sep=4pt,font=\scriptsize,minimum height=14mm,text width=36mm,fill=white},
 sy/.style={bx,draw=kunc,line width=0.9pt,fill=kunc!6},
 nu/.style={bx,draw=limc,line width=0.9pt,fill=limc!5},
 res/.style={bx,draw=lic,line width=1.0pt,fill=lic!6,text width=30mm},
 aux/.style={draw=black!40,dashed,rounded corners=2pt,align=center,execute at begin node=\nohyII,
       inner sep=3pt,font=\scriptsize,text width=36mm,fill=black!2},
 lab/.style={font=\small\bfseries,anchor=east,align=right},
 ar/.style={->,black!65},
 cmp/.style={<->,black!50,dashed}]

\node[font=\small] at (0,1.35) {grade decomposition};
\node[font=\small] at (4.3,1.35) {1st stage ($\M_1$)};
\node[font=\small] at (8.6,1.35) {2nd stage ($\M_2$)};

\node[lab,text=kunc] at (-2.15,0) {symbolic\\[-1pt]\scriptsize (Mathematica)};
\node[sy] (s1) at (0,0) {ADM Hamiltonian in scalar invariants;\\
  $\bm P\to\tau\bm P$, $\bm R\to\bm R/\tau^2$, Legendre series of $1/r_{a3}$ to $k=5$;\\
  coefficient of $\tau^m$ = order $(1,m/2)$ (Table~\ref{tab:grades})};
\node[sy] (s2) at (4.3,0) {inner averages as Laurent polynomials in $u=1-e_1\cos E$ with
  $\av{u^k}_E$; brackets: inner Delaunay $+$ outer Cartesian, $\tau$ scaling of the
  orders};
\node[sy] (s3) at (8.6,0) {$w_2$-weighted trigonometric polynomials in $F$;
  antiderivatives and averages as linear maps; $v=F-\M_2$ by Eq.~\eqref{eq:vh};
  exact pairs at fixed $J$};

\node[lab,text=limc] at (-2.15,-2.6) {numerical\\[-1pt]\scriptsize (Python, FFT)};
\node[nu] (n1) at (0,-2.6) {Hamiltonian at complex $\tau$ on a circle, $1/r_{a3}$ kept
  analytic; grades extracted by FFT (discrete Cauchy integral)};
\node[nu] (n2) at (4.3,-2.6) {antiderivatives in $\M_1$ by FFT (512 points);
  brackets by central differences};
\node[nu] (n3) at (8.6,-2.6) {antiderivatives in $\M_2$ by FFT (128 points);
  harmonics in $(\omega_1,\omega_2)$ by 2D FFT; exact reduced pairs at fixed $J$};

\node[res] (K) at (12.6,-1.3) {secular Hamiltonian\\ $K^{(1,3)},\,K^{(1,7/2)}$,\\ $K^{(1,4)},\,K^{(0,9/2)}$\\
  (Secs.~\ref{sec:mono}--\ref{sec:q2}, App.~\ref{app:expr})};
\node[aux,text width=56mm] (chk) at (2.15,-4.65) {other checks: EIH $\equiv$ ADM; osculating--contact
  offsets; PPN extension; gauge generator $\gamma_\tau$ and $\{K_{\rm Q},\gamma_\tau\}$};
\node[aux,text width=44mm] (code) at (10.6,-4.65) {C++ secular code of Paper~I with the
  canonical terms added (gradient by central differences) $\to$ Sec.~\ref{sec:orbit}};

\draw[ar] (s1) -- (s2); \draw[ar] (s2) -- (s3);
\draw[ar] (n1) -- (n2); \draw[ar] (n2) -- (n3);
\draw[ar] (s3.east) -- ++(0.55,0) |- ([yshift=2mm]K.west);
\draw[ar] (n3.east) -- ++(0.55,0) |- ([yshift=-2mm]K.west);
\draw[cmp] (s1) -- node[right,font=\scriptsize,text=black!60]{same grades} (n1);
\draw[cmp] (s2) -- node[right,font=\scriptsize,text=black!60]{agree to $10^{-9}$} (n2);
\draw[cmp] (s3) -- node[right,font=\scriptsize,text=black!60]{agree} (n3);
\draw[ar] (K.south) -- (code.north -| K.south);
\end{tikzpicture}}
\caption{Symbolic (blue) and numerical (red) pipelines used to construct the secular terms.
Both start from the 1PN ADM Hamiltonian, separate the orders by the scaling parameter
$\tau$, and carry out the two Lie--Hori--Deprit stages of Sec.~\ref{sec:lie}; they share no
code and are compared at each stage (dashed double arrows). The resulting secular
Hamiltonian (green) is implemented in the C++ secular code used in Sec.~\ref{sec:orbit}.
The independent checks listed at the end of this Appendix are shown in the dashed box.}
\label{fig:pipe}
\end{figure*}

The two pipelines are summarized in Fig.~\ref{fig:pipe}.

\paragraph*{Grade decomposition.}
Eq.~\eqref{eq:adm} is written in scalar invariants of the Jacobi variables
($r$, $R$, $\hat{\bm n}\!\cdot\!\hat{\bm N}$ and the products of $\bm r,\bm R,\bm p,\bm P$),
the momenta and distances are replaced by $\bm P\to\tau\bm P$ and $\bm R\to\bm R/\tau^2$,
$1/r_{a3}$ is expanded in Legendre polynomials up to $k=5$, and the coefficient of $\tau^m$
is the part of order $\delta\varepsilon^{m/2}$. Table~\ref{tab:grades} lists the number of
terms and the parities under $\Pi_1:(\bm r,\bm p)\to-(\bm r,\bm p)$ (equivalent to
$m_1\leftrightarrow m_2$ and $\omega_1\to\omega_1+\pi$) and
$\Pi_2:(\bm R,\bm P)\to-(\bm R,\bm P)$ ($\omega_2\to\omega_2+\pi$). Both are canonical,
so that parities multiply in Poisson brackets. The part of order $\delta\varepsilon^{3/2}$
has no terms at all.

\begin{table}[tb]
\caption{Number of terms of the 1PN Hamiltonian at each order in scalar invariants of the
Jacobi variables, and parities $(\Pi_1,\Pi_2)$ ($\pm$ mixed $=0$). The Newtonian part has
1, 2, 4, 4, 6, 6 terms at $(0,0)$, $(0,1)$, $(0,3)$, $(0,4)$, $(0,5)$, $(0,6)$.}
\label{tab:grades}
\centering\small
\begin{tabular}{@{}lcccccccccc@{}}
\toprule
$m/2$ & $0$ & $\frac12$ & $1$ & $\frac32$ & $2$ & $\frac52$ & $3$ & $\frac72$ & $4$ & $\frac92$\\
\midrule
terms & 8 & 6 & 9 & \textbf{0} & 20 & 10 & 8 & 18 & 38 & 22\\
$\Pi_1$ & $+$ & $-$ & $+$ & & $0$ & $+$ & $+$ & $-$ & $0$ & $+$\\
$\Pi_2$ & $+$ & $-$ & $+$ & & $0$ & $+$ & $+$ & $-$ & $0$ & $+$\\
\bottomrule
\end{tabular}
\end{table}

\paragraph*{Averages.}
Inner averages are taken in the eccentric anomaly by writing the integrand as a Laurent
polynomial in $u=1-e_1\cos E$ and using the table of $\av{u^k}_E$; negative powers of
$u$ and of $1+e_2\cos F$ are carried as separate positive symbols so that automatic
expansion does not destroy the structure. Outer integrands multiplied by $w_2$ are
trigonometric polynomials in $F$, stored as dictionaries of monomials
$\cos^aF\sin^bF$; antiderivatives and averages are linear maps on these dictionaries.
Terms containing the equation of the center $v$ are reduced by Eq.~\eqref{eq:vh}.

\paragraph*{Independent numerical construction.}
For the next-to-leading order 1PN\,$\times$\,quadrupole term and the magnetic-quadrupole split we
built an independent Python (NumPy/SciPy) implementation. The ADM Hamiltonian is evaluated
at complex $\tau$ on a circle, with
$1/r_{a3}=\tau^2/\sqrt{(\bm R-\tau^2\bm x_a)\cdot(\bm R-\tau^2\bm x_a)}$ so that it remains
analytic, and the grades are extracted by FFT (a discrete Cauchy integral). Antiderivatives
in $\M_1$ and $\M_2$ are taken by FFT (512 and 128 points), brackets by central
differences; the second stage uses the exact reduced pairs at fixed $J$. The quadrupole-squared
term was evaluated in the same way, and its harmonics by a two-dimensional FFT in
$(\omega_1,\omega_2)$.

\paragraph*{Other checks.}
(i) The identity $H_{\rm 1PN}=-L_{\rm 1PN}(\bm x,\bm p/m)$ between Eqs.~\eqref{eq:LEIH}
and \eqref{eq:adm} was verified symbolically for three bodies with generic vectors.
(ii) The osculating--contact offsets, Eqs.~\eqref{eq:oscon-a}--\eqref{eq:oscon-m3}, were computed
symbolically from $\bm v-\bm p/\mu_1=\partial H_{\rm 1PN}/\partial\bm p$ and checked by direct
numerical averaging. (iii) The PPN extension, Eq.~\eqref{eq:h11ppn}, and the two-body periastron
advance $(2+2\gamma-\beta)/3$ of the same Lagrangian were checked numerically to twelve
digits. (iv) The gauge generator, Eq.~\eqref{eq:gamclosed}, and its bracket with $K_{\rm Q}$ were
derived symbolically and checked against the numerical second-order calculation.
The symbolic calculations were carried out with Mathematica, the numerical ones with
Python, and the orbital integrations of Sec.~\ref{sec:orbit} with the C++ secular code of
Paper~I, to which the canonical terms of this paper were added.

\section{Intermediate expressions}
\label{app:expr}

\paragraph*{Newtonian quadrupole: inner average and first-stage generator.}
The quadrupole part of Eq.~\eqref{eq:HN} is
$H^{(0,3)}=-(G\mu_1m_3/2R^5)\big[3(\bm r\!\cdot\!\bm R)^2-r^2R^2\big]$. In the components
$(X,Y,Z)$ of $\bm R$ along the inner pericenter, in-plane normal and orbit normal
$(\hat{\bm\alpha},\hat{\bm\beta},\hat{\bm\gamma})$, with $e=e_1$ and $\ell=\ell_1$, its inner
average is
\begin{equation}
\begin{split}
\av{H^{(0,3)}}_{\M_1}={}&-\frac{G\mu_1m_3a_1^2}{4R^5}
\big[(1+9e^2)X^2+(1-6e^2)Y^2\\
&-(2+3e^2)Z^2\big],
\end{split}
\label{eq:HQavg}
\end{equation}
which gives $K_{\rm Q}=A_{\rm Q}Z$, Eq.~\eqref{eq:KQ}, after the outer average with
$\av{R^iR^j/R^5}_{\M_2}$ (Sec.~\ref{sec:pnq-e}). The time-mean-free solution of the
homological equation, Eq.~\eqref{eq:homo}, which enters every first-stage cross term
$a\otimes3$, is a trigonometric polynomial of degree three in the eccentric anomaly,
\begin{equation}
\begin{split}
n_1W^{(0,3)}={}&-\frac{G\mu_1m_3a_1^2}{2R^5}\big[X^2S_X+Y^2S_Y\\
&+Z^2S_Z+XY\,C_{XY}\big],
\end{split}
\label{eq:WQ}
\end{equation}
with
\begin{align}
S_X&=\tfrac14e(9e^2-19)\sin E+\tfrac34(1+e^2)\sin2E\nonumber\\
&\quad+\tfrac1{12}e(e^2-3)\sin3E,\nonumber\\
S_Y&=\tfrac14e(11-6e^2)\sin E-\tfrac34\sin2E\nonumber\\
&\quad+\tfrac1{12}e(3-2e^2)\sin3E,\nonumber\\
S_Z&=\tfrac14e(8-3e^2)\sin E-\tfrac34e^2\sin2E+\tfrac1{12}e^3\sin3E,\nonumber\\
C_{XY}&=\ell\big[\tfrac{15}4e^2+\tfrac{15}2e\cos E-\tfrac32(1+e^2)\cos2E\nonumber\\
&\quad+\tfrac12e\cos3E\big].
\label{eq:WQcoef}
\end{align}
The constant $\tfrac{15}{4}e^2\ell$ in $C_{XY}$ is the integration constant that makes
$\av{W^{(0,3)}}_{\M_1}=0$ with the time measure; with another measure it would change by a
function of the slow variables, which is the freedom $\gamma$ of Eq.~\eqref{eq:homo}. For
$e\to0$ only the second harmonic survives,
$n_1W^{(0,3)}\to-(3G\mu_1m_3a_1^2/8R^5)\big[(X^2-Y^2)\sin2E-2XY\cos2E\big]$.
Eqs.~\eqref{eq:WQ}--\eqref{eq:WQcoef} were checked symbolically against
$n_1\partial_{\M_1}W^{(0,3)}=H^{(0,3)}-\av{H^{(0,3)}}_{\M_1}$.

\paragraph*{Vector form of the canonical terms.}
For the orbit integrations it is convenient to write the inner-orbit terms in the vectors
$\bm j=\ell_1\hat{\bm\gamma}$ and $\bm e=e_1\hat{\bm\alpha}$ and the outer basis
$(\hat{\bm u}_2,\hat{\bm v}_2,\hat{\bm z}_2)$. With $j_z=\bm j\!\cdot\!\hat{\bm z}_2$ and
$e_z=\bm e\!\cdot\!\hat{\bm z}_2$, the electric 1PN\,$\times$\,quadrupole term,
Eq.~\eqref{eq:K13}, reads
\begin{equation}
K^{(1,3)}=U_3\bigg[F_0\Big(1-\frac{3j_z^2}{j^2}\Big)
+\frac{F_2}{e_1^2}\Big(e_1^2-\frac{e_1^2j_z^2}{j^2}-2e_z^2\Big)\bigg],
\label{eq:K13vec}
\end{equation}
with $U_3=G^2m_1m_2m_3a_1/(c^2a_2^3\ell_2^3)$; $F_2/e_1^2$ is regular at $e_1=0$, where it
equals $9(4-\nu)/32$. The next-to-leading order term, Eq.~\eqref{eq:K14}, is already in vector form
through the second lines of Eqs.~\eqref{eq:Z} and \eqref{eq:Y}. For a secular Hamiltonian
$K(\bm j,\bm e)$ of the inner orbit the equations of motion are the Milankovitch equations
\cite{Milankovitch1939,AllanWard1963,RosengrenScheeres2014}
\begin{align}
\frac{d\bm j}{dt}&=-\frac1{L_1}\big(\bm j\times\nabla_{\bm j}K+\bm e\times\nabla_{\bm e}K\big),
\nonumber\\
\frac{d\bm e}{dt}&=-\frac1{L_1}\big(\bm j\times\nabla_{\bm e}K+\bm e\times\nabla_{\bm j}K\big),
\label{eq:milank}
\end{align}
with $L_1=\mu_1\sqrt{Gma_1}$; this is how the canonical terms enter the code of
Sec.~\ref{sec:orbit}.

\paragraph*{Electric 1PN\,$\times$\,quadrupole term before the outer average.}
In the components $(X,Y,Z)$ of $\bm R$ along $(\hat{\bm\alpha},\hat{\bm\beta},\hat{\bm\gamma})$,
the inner averages of the direct term and of the first-stage cross term are
\begin{widetext}
\begin{align}
\av{H^{(1,3)}}_{\M_1}&=\frac{G^2m_1m_2m_3\,a_1}{8c^2m^2R^5}\Big\{\nonumber\\
&\hspace{2em}(m_1^2+m_2^2)\big[(7e_1^2-1)X^2-(1+2e_1^2)Y^2+(2-5e_1^2)Z^2\big]\nonumber\\
&\hspace{2em}+3m_1m_2\big[(1+e_1^2)X^2+(1-2e_1^2)Y^2+(e_1^2-2)Z^2\big]\Big\},\\
\av{\{\Delta^{(1,0)},W^{(0,3)}\}}_{\M_1}
&=\frac{G^2m_1m_2m_3\,a_1}{8c^2e_1^2\ell_1R^5}\Big\{12(1-\ell_1)(X^2-Y^2)\nonumber\\
&\hspace{2em}+2e_1^2\big[(26\ell_1-42)R^2+(81-60\ell_1)X^2+9(5-2\ell_1)Y^2\big]\nonumber\\
&\hspace{2em}+e_1^4\big[4(2\ell_1-9)R^2+(81-48\ell_1)X^2+3(9+8\ell_1)Y^2\big]\Big\}.
\end{align}
\end{widetext}
The cross term depends on the masses only through $m_1m_2m_3$. It was checked against the
numerical evaluation (512 points in $\M_1$) to $10^{-9}$ and against the reversed bracket
$\av{\{\Delta^{(0,3)},W^{(1,0)}\}}$.

\paragraph*{Next-to-leading order 1PN\,$\times$\,quadrupole term: contributions.}
With
\begin{align*}
Q&=2(1+5e_2^2)m^2+(23e_2^2-2)mm_3-6(1-e_2^2)m_3^2,\\
Q_2&=4(1+4e_2^2+3\ell_2)(m^2+m_3^2)\\
&\quad+(8+27e_2^2+24\ell_2)mm_3,\\
Y_2&=8(e_2^2-1)(e_2^2-2+2\ell_2)(m^2+m_3^2)\\
&\quad+\big[11e_2^4+32(1-\ell_2)+16e_2^2(2\ell_2-3)\big]mm_3,\\
Y_s&=4(1-\ell_2)+e_2^2(6\ell_2-8)+e_2^4(4+3\ell_2),
\end{align*}
the contributions to Eqs.~\eqref{eq:cZ}--\eqref{eq:cY} are those of Table~\ref{tab:cZcY}.
All three $c_Y$ contributions are $\mathcal{O}(e_2^2)$ for $e_2\to0$.

\begin{table}[tb]
\caption{Contributions to the coefficients $c_Z$ and $c_Y$ of the next-to-leading order 1PN\,$\times$\,quadrupole term, Eqs.~\eqref{eq:cZ}--\eqref{eq:cY}. ``Outer 1PN $\otimes$ ZLK'' and ``de Sitter $\otimes$ ZLK'' denote second-stage cross terms, in the sense of Sec.~\ref{sec:lie}, of the secular outer 1PN and de Sitter Hamiltonians with the ZLK Hamiltonian.}
\label{tab:cZcY}
\centering\small
\begin{tabular}{@{}lll@{}}
\toprule
Contribution & $c_Z$ & $c_Y$\\
\midrule
direct $+$ 1st stage & $-\dfrac{Q}{64mM}$ & $-\dfrac{3e_2^2(4m+7m_3)}{128M}$\\[2mm]
outer 1PN $\otimes$ ZLK & $-\dfrac{3Q_2}{64mM}$ & $-\dfrac{3Y_2}{128e_2^2mM}$\\[2mm]
de Sitter $\otimes$ ZLK & $0$ & $-\dfrac{m_3(4m+3m_3)Y_s}{64e_2^2\ell_2mM}$\\
\bottomrule
\end{tabular}
\end{table}

\paragraph*{Octupole-type coefficients.}
With $h=\ell_1$ and $e=e_1$,
\begin{widetext}
\begin{align}
c_3&=-\frac{45\,\mathcal P_3}{1024e^3h}-\frac{675}{512}e^3m_1m_2+\frac{75}{512}e^3(5m_1^2+8m_1m_2+5m_2^2),\nonumber\\
c_1&=-\frac{3\,\mathcal P_1}{2048eh}-\frac{9}{1024}e(4+11e^2)m_1m_2
+\frac{3}{1024}e\big[(29e^2-4)(m_1^2+m_2^2)+8(4+e^2)m_1m_2\big],\nonumber\\
\mathcal P_3&=\big[32(1-h)+8e^2(17h-19)+8e^4(17h-9)+e^6(200h-81)\big](m_1^2+m_2^2)\nonumber\\
&\quad+2\big[32(1-h)+8e^2(17h-19)+8e^4(17h-9)+3e^6(65h-27)\big]m_1m_2,\nonumber\\
\mathcal P_1&=\big[216(h-1)+20e^2(46h-81)+e^4(184h-495)\big](m_1^2+m_2^2)\nonumber\\
&\quad+2\big[216(h-1)+4e^2(229h-405)+e^4(173h-495)\big]m_1m_2,
\end{align}
\end{widetext}
with the expansions
\begin{align*}
c_3&=-\tfrac{1}{512}e^3\big[435(m_1^2+m_2^2)+1470m_1m_2\big]+\mathcal{O}(e^5),\\
c_1&=\tfrac{3}{64}e\big[25(m_1^2+m_2^2)+52m_1m_2\big]+\mathcal{O}(e^3).
\end{align*}

\bibliographystyle{apsrev4-2}
\bibliography{refsII}

\end{document}